\documentclass[useAMS,usenatbib]{mn2e}
\usepackage{times}
\usepackage{natbib}
\usepackage{aas_macros}
\usepackage{amsmath}
\usepackage[pdftex]{graphicx}

\title[]{Radio Halos From Simulations And Hadronic Models II: The Scaling Relations of Radio Halos}
\author[J. Donnert, K. Dolag, R.Cassano, G. Brunetti]{
J. Donnert$^{1,2}$,\thanks{jdonnert@mpa-garching.mpg.de}
K. Dolag$^{1}$,  
R. Cassano$^{2}$, 
G. Brunetti$^{2}$\\
$1$Max Planck Institute for Astrophysics, P.O. Box 1317, D--85741 Garching, Germany\\
$2$INAF Istituto di Radioastronomia, via P. Gobetti 101, I-40129 Bologna, Italy\\
}

\begin{document}

\date{Accepted ???. Received ???; in original form ???}

\pagerange{\pageref{firstpage}--\pageref{lastpage}} \pubyear{2008}

\maketitle

\label{firstpage}

\begin{abstract}
We use results from a constrained, cosmological MHD simulation 
of the Local Universe to predict radio haloes and their evolution
for a volume limited set of galaxy clusters and compare to current 
observations. The simulated magnetic field inside the clusters 
is a result of turbulent amplification within them, with 
the magnetic seed originating from star-burst driven, galactic outflows. 
We evaluate three models, where we choose different normalisations 
for the cosmic 
ray proton population within clusters. Similar to our previous analysis  of the Coma cluster \citep{2010MNRAS.401...47D},
the radial profile and the morphological properties of observed radio halos 
can not be reproduced, even with a radially increasing energy fraction within
the cosmic ray proton population. Scaling relations between X-ray 
luminosity and radio power can be reproduced by all models, however 
all models fail in the prediction of clusters with no radio emission.
Also the evolutionary tracks of our largest clusters in all models
fail to reproduce the observed bi-modality in radio luminosity. 
This provides additional evidence that the framework of hadronic, 
secondary models is disfavored to reproduce the large scale diffuse 
radio emission.
of galaxy clusters. We also provide predictions for the unavoidable 
emission of $\gamma$-rays from the hadronic models for the full cluster set. 
None of such secondary models is  yet excluded by the observed 
limits in $\gamma$-ray emission, emphasizing that large scale diffuse radio 
emission is a powerful tool to constrain the amount of cosmic ray protons 
in galaxy clusters.
\end{abstract}

\begin{keywords}
galaxies:clusters
\end{keywords}

\section{Introduction}\label{intro}

The thermal gas, that is the dominant component in the
Inter-Galactic-Medium (IGM), is mixed with
magnetic fields and relativistic particles, as proven
by radio observations which detected Mpc-sized diffuse radio 
emission from the IGM, in the form of radio halos and relics
\citep[e.g.][]{2003astro.ph..1576F,2008SSRv..134...93F}.
These Mpc-scale radio sources are found in a fraction of massive
clusters with complex dynamics, which suggests a connection between
non-thermal emission and cluster mergers 
\citep[e.g.][]{2001ApJ...553L..15B,2008A&A...484..327V,2009A&A...507..661B}
Cluster mergers are the most energetic events in the universe and 
a fraction of the energy dissipated during these
mergers may be channelled into the amplification of the
magnetic fields \citep[e.g.][]{2002A&A...387..383D,2006MNRAS.366.1437S,2008Sci...320..909R}
and into the acceleration of relativistic, primary,
electrons and protons via shocks and turbulence 
\citep[e.g.][]{1998A&A...332..395E,1999astro.ph.11439S,2001MNRAS.320..365B,
2004MNRAS.350.1174B,2001ApJ...557..560P,2003ApJ...583..695G,2003ApJ...593..599R,
2005MNRAS.357.1313C,2006MNRAS.367..113P,2007MNRAS.378..245B,2009A&A...504...33V}

Relativistic protons in the IGM have long
life-times and remain confined within galaxy clusters for a Hubble
time \citep[e.g.][]{1996SSRv...75..279V,1997ApJ...487..529B}. 
As a consequence
they are expected to be the dominant non-thermal particle component.
Collisions between these relativistic protons and the thermal
protons in the IGM generate secondary particles that combined with 
the primary relativistic particles are expected to produce a complex 
emission spectrum from radio to $\gamma$-rays
\citep[e.g.][]{2001APh....15..223B,2009A&A...507..661B}.
Only upper limits to the $\gamma$-ray emission from galaxy clusters have
been obtained so far 
\citep[][]{2003ApJ...588..155R,2006ApJ...644..148P,
2009A&A...495...27A,2009A&A...502..437A,2009arXiv0909.3267T} 
however the FERMI Gamma-ray
telescope will shortly allow a step forward, having a chance to obtain
first detections of galaxy clusters or to put stringent constraints
on the energy density of the relativistic protons.
Most importantly, in a few years the Low Frequency Array (LOFAR) and the 
Long Wavelength Array (LWA) will observe galaxy clusters at low radio 
frequencies with the potential to discover the bulk of the 
cluster-scale synchrotron emission in the Universe 
\citep[e.g.][]{2002A&A...396...83E,2009arXiv0910.2025C,2006MNRAS.369.1577C}.

The emerging theoretical picture is very complex and modern 
numerical simulations provide an efficient way to obtain 
detailed models of non thermal emission from galaxy clusters to
compare with present and future observations.
Advances in this respect have been recently obtained by including
aspects of cosmic-ray physics into cosmological Lagrangian simulations 
mostly focussing on the acceleration of relativistic particles at shocks 
and on the relative production of secondary electrons 
\citep[e.g.][]{2008MNRAS.385.1211P}.
In this work we investigate the non-thermal emission from secondary
particles in galaxy clusters extracted from Lagrangian
cosmological simulations and, for the first time, we report on an 
adequate comparison between our expectations and observations.

\section{Simulations}\label{sims}
The simulation was done using the cosmological simulation code
GADGET-2 \citep{2005MNRAS.364.1105S} with a treatment for magnetic
fields. It features an entropy conserving formulation of Smooth
Particle Hydrodynamics (SPH) \citep{2002MNRAS.333..649S}, which is
supplemented with the formulation of ideal MHD presented in
\citet{2008arXiv0807.3553D}. The implementation follows the induction
equation and computes the back reaction of the magnetic field using a
symmetric formulation of the Lorentz force. We used a divergence
cleaning scheme presented in \citet{2001ApJ...561...82B}, which
reduces numerical noise in shocks by subtracting the magnetic force
which is proportional to the divergence of the field. It also helps to
suppress the clumping instability particle based MHD codes encounter
in regions with small plasma $\beta$ (i.e. where magnetic pressure
considerably exceeds thermal pressure). \par In non radiative
simulations like ours, regions with small plasma $\beta$ are
rare. Only the strong shocks in cores of galaxy clusters during major
mergers produce enough compression to amplify the field
to become dynamically dominant. 
These mergers are relatively brief events and are
handled more accurately with our new numerical treatment 
\citep[see ][ for details]{2008arXiv0807.3553D}. \\ 
\citet{2006MNRAS.367.1641B} have shown that non radiative
simulations overpredict the gas density in cores of galaxy
clusters. This affects our simulation as well and can be seen in density
and magnetic field profiles as well as in X-ray luminosities. As
cosmological MHD SPH simulations lack physical dissipation, 
radiative SPH MHD simulations are not feasible at the moment.
On the other hand secondary models have difficulties reproducing the outer 
parts of radio haloes correctly. Therefore our main focus lies on these regions, 
where the simulations are not affected by the overpredicted gas density.

\subsection{Initial Conditions}\label{inicond}
We used a constrained realisation of the local universe (see
\citet{2005JCAP...01..009D} and references therein). The initial
conditions are similar to those used in \citet{2002MNRAS.333..739M} to
study the formation of the local galaxy population. They were obtained
based on the IRAS 1.2-Jy galaxy survey. Its density field was
smoothed on a scale of $7\, \mathrm{Mpc}$, evolved back in time to
$z=50$ using the Zeldovich approximation and assumed to be
Gaussian \citep{Hoffman1991}. The IRAS observations constrain a volume
of $\approx 115 \, \mathrm{Mpc}$ centered on the Milky Way. It was
sampled with dark matter particles and embedded in a periodic box of
$\approx 343 \, \mathrm{Mpc}$ comoving. Outside of the inner region,
the box is filled with dark matter particles with $1/6$th of the
resolution, to cover for long range gravitational tidal forces arising
from the low-frequency constrains. \\ 
In the evolved density field, many locally observed galaxy clusters can 
be identified by position and mass. Especially the Coma cluster 
\citep[see][]{2010MNRAS.401...47D} shows
remarkable similarities in morphology. A fly-through of the simulation
can be downloaded from the MPA
Website\footnote{http://www.mpa-garching.mpg.de/galform/data\_vis/index.shtml\#movie12}.\\ The
initial conditions were extended to include gas by splitting dark
matter particles in the high resolution region into gas and dark
matter particles of masses $0.69 \times 10^9\; {\rm M}_\odot$ and $4.4
\times 10^9\; {\rm M}_\odot$ respectively. Therefore the biggest
clusters are resolved by about a million particles. The gravitational
softening length was set to $10\,\mathrm{kpc}$. This is comparable to
the inter-particle separation found in the centre of the largest
clusters. 

\subsection{Magnetic Fields from Galactic Outflows}\label{wind}

\begin{figure*}
\includegraphics[width=0.9\textwidth]{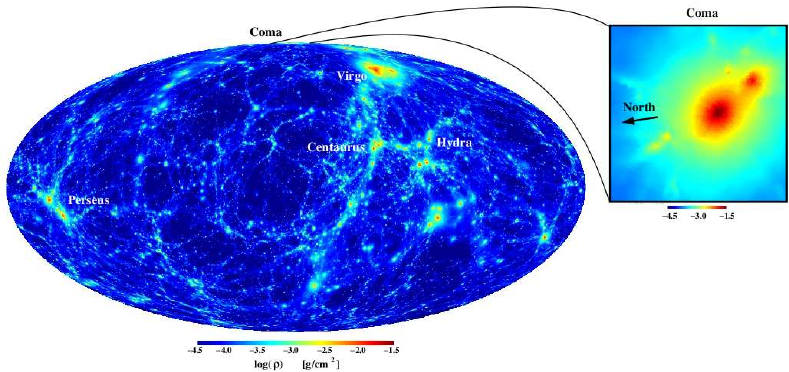}
\includegraphics[width=0.9\textwidth]{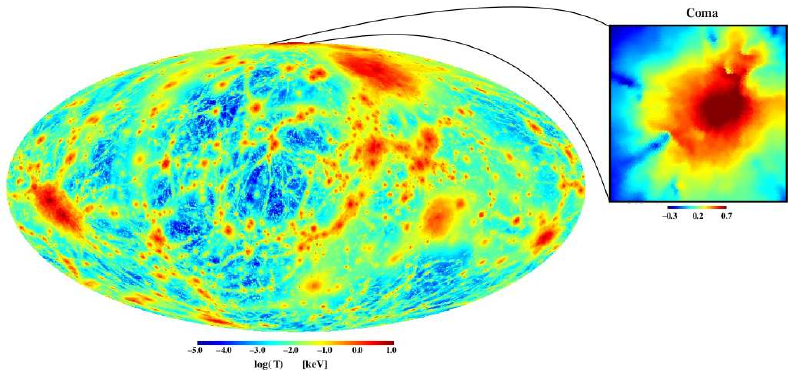}
\includegraphics[width=0.9\textwidth]{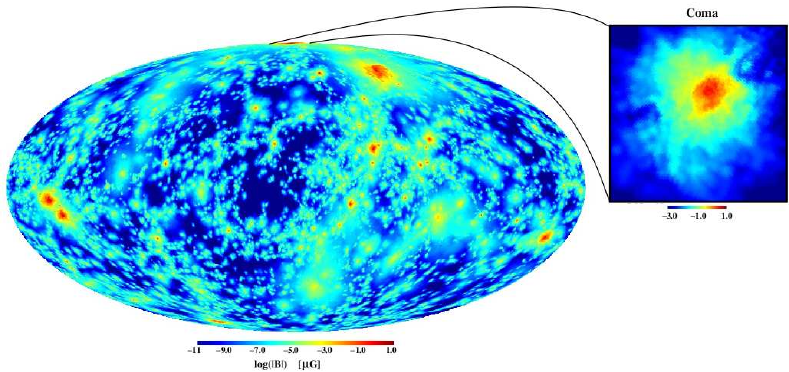}
\caption{{Full sky maps of the simulation in galactic cordinates. 
From top to bottom, electron density, temperature and 
magnetic field, projected through the whole box}. 
The inlay shows a zoom onto a 3$\times$3 degree 
region arround the Coma cluster, respectively. In the upper most 
map, the most prominent clusters of the local universe are labled
and the arrow in the inlay points towards north.}\label{vis}
\end{figure*}

The origins of magnetic fields in galaxy clusters are still under
debate. It is assumed that some kind of early seed magnetic field is
amplified by structure formation through adiabatic compression,
turbulence and shear flows to values observed today ($\approx 1 - 10
\,\mu\mathrm{G}$ in clusters).  Three main classes of models for the
seed field exist: At first the seed fields can be created in shocks
through the ''Biermann battery ''
\citep{1997ApJ...480..481K,Ryu..1998,2001ApJ...562..233M}. A second
class of models invokes primordial processes to predict seed fields
that fill the entire volume of the universe. The coherence length of
these fields strongly depends on the details of the model \citep[see
][ for a review]{Grasso..PhysRep.2000}. Finally the seed can be
produced by AGN \citep{1997ApJ...477..560E,2001ApJ...556..619F} or
starbursting galaxies \citep{Volk&Atoyan..ApJ.2000} at high redshift
($z \approx 4 - 6$), whose outflows contaminate the proto-cluster
region.\par  Cosmological simulations using SPH
\citep{1999A&A...348..351D,2002A&A...387..383D,2005JCAP...01..009D}
and grid based Adaptive Mesh Refinement (AMR) codes
\citep{2005ApJ...631L..21B,2008A&A...482L..13D,2008ApJS..174....1L}
were able to show that observed Faraday rotations are compatible with
a cosmological seed field of $\approx 10^{-11}\,\mathrm{G}$. They also
suggest that spatial distribution and structure of cluster magnetic
fields are determined by the dynamics in the velocity field caused by
structure formation
\citep{1999A&A...348..351D,2002A&A...387..383D}.\par For this work we
follow \citet{2008arXiv0808.0919D} in terms of magnetic field
origin. They use a semianalytic model for galactic winds
\citep{2006MNRAS.370..319B} to seed magnetic fields in a constrained
cosmological MHD SPH simulation. The continuous seeding process is
approximated with an instantaneous seed at $z\approx 4$. As they were
able to show, the main properties of magnetic fields obtained in
clusters were not influenced by that approximation. \\
\begin{table}
    \centering
    \begin{tabular}{c|c|c} \hline
        Parameter & Value & Source \\ \hline
        $R_{0}$           & $400\,\mathrm{pc}$ 			& \citep{1988AA...190...41K}  \\
        $B_{0}$ 			  & $5\,\mu \mathrm{G}$      	&  \citep{2008arXiv0808.0919D} \\
        $B_{\mathrm{G}}$  & $3\,\mu\mathrm{G}$ 			&   \citep{2008arXiv0808.0919D} \\
        $\dot{M}_{\star}$ & $10\,\mathrm{M}_{\odot}/{\mathrm{yr}}$ & \citep{2001astro.ph..6564D} \\
        $t_{\mathrm{sb}}$ & $ 150\,\mathrm{Myr}$      & \citep{2001astro.ph..6564D} \\
        $M_{\mathrm{ISM}}$& $<300 \times 10^{12} \mathrm{M}_{\odot}$  & from simulation \\
		  \hline
    \end{tabular}
\caption{Summary of the parameters used for the wind model. This
  corresponds to the {\it 0.1 Dipole} simulation in
  \citep{2008arXiv0808.0919D}, which fits best to observations of
  Faraday rotation}. \label{param_table}
\end{table}
The wind model used assumes adiabatic expansion of a spherical gas bubble
with homogeneous magnetic energy density around every galaxy below a
certain mass threshold. The magnetic bubble can be characterised by
radius and field strength. The galaxy injects gas into the bubble
carring frozen-in magnetic field from the disc into the bubble over
the star-burst timescale. Its final size is determined by the wind
velocity, which is a function of the star formation rate and the
properties of the ISM. \citet{2006MNRAS.370..319B} give an evolution
equation for the magnetic energy in the bubble depending on the
star-burst timescale. The energy is converted into a dipole
moment and seeded once at a chosen redshift. The magnetic field is
then amplified by structure formation to $\mu \mathrm{G}$ level. For
details on the wind model refer to \citet{2006MNRAS.370..319B,2008arXiv0808.0919D}. 
Figure \ref{vis} shows full sky maps produced from the simulation, 
projecting the electron density, temperature and the magnetic field. 
The magnetic field closely follows the density distribution. 
The magnetic field is more patchy  in the filaments compared to a cosmological 
seed because the seeding by individual galaxies does not overlap.

\par

\begin{figure}
\centering
\includegraphics[width=0.45\textwidth]{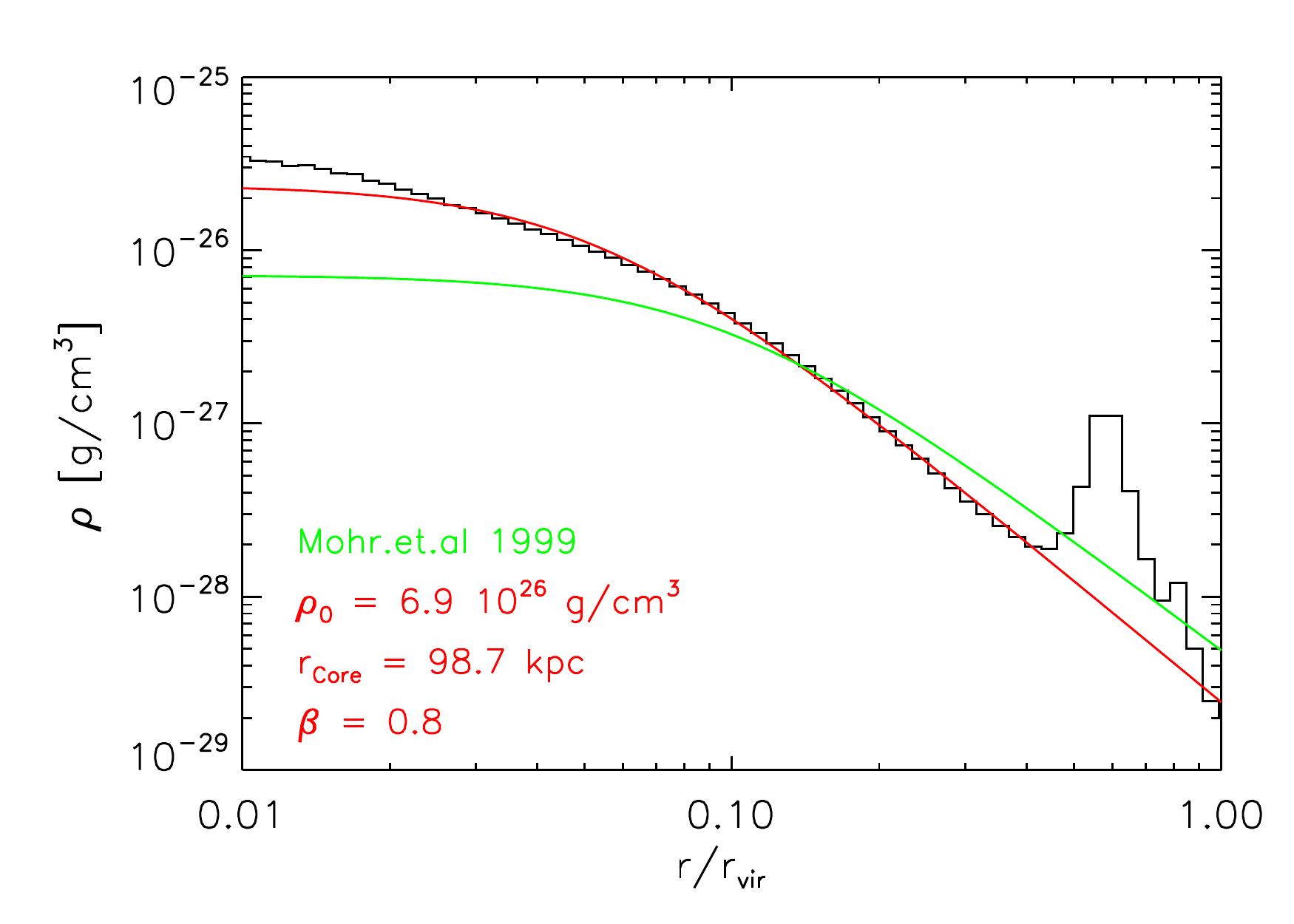}
\caption{Radial profile of the density in our simulated Coma
  cluster. In red we plot the best fit beta model
  ($\rho_{0}=6.4\times10^{15}, \beta= 0.8, r_{\mathrm{core}}=0.06$),
  excluding the density bump at 0.6 $R_{\mathrm{vir}}$. A fit to
  observations of Coma by \citet{1999ApJ...517..627M} is shown in
  green. Due to the non-radiativity of our simulation the gas density
  is overestimated by a factor of 5 in the center}\label{rho_bmodel}
\end{figure}

The simulation used in this work is based on the {\it 0.1 Dipole}
parameter set from \citet{2008arXiv0808.0919D}, shown in table
\ref{param_table}. It represents the best fit to observations of
Faraday rotation presented in \citet{2008arXiv0808.0919D}. There the
parameter space was explored based on observations from the wind in
M82. It was demonstrated that the resulting magnetic field does not
critically depend on the exact choice of the parameters.

\section{Modelling Hadronic Secondary Electrons in Clusters}\label{CRpop}

In secondary models proposed for the origin of radio halos
\citep{1980ApJ...239L..93D,1999APh....12..169B},
the emitting cosmic ray electrons (CRe) are a product of cosmic ray proton
collisions (CRp) with thermal protons in the cluster. 
Possible mechanisms for the injection of CRp in the IGM 
include shock waves caused by cluster accretion and mergers
\citep[e.g.][]{2003ApJ...583..695G,2003ApJ...593..599R,
2006MNRAS.367..113P,2009A&A...504...33V}
outflows from radio galaxies (AGN) \citep[e.g.][]{1998A&A...332..395E,2008arXiv0808.0349R}
and supernova driven winds
\citep{1996SSRv...75..279V}.
The resulting CRp spectrum from these theoretical scenarios is
a power law \citep[e.g.][]{2002cra..book.....S},  
that is also consistent with observations of
galactic cosmic rays.

We assume a power law spectrum of CRp in the IGM and calculate 
the spectrum of high-energy secondary electrons 
under stationary conditions by considering only synchrotron and 
inverse Compton losses and by neglecting re-acceleration 
processes in the IGM (Section \ref{theory}).

\noindent
Throughout the first two of our models,
we use the magnetic fields obtained from the constrained cosmological
simulation described in section \ref{wind}. In a third model we 
increased the magnetic field, especially at larger distances to the center, by up-scaling
the magnetic field in our simulation to be $B \propto \sqrt{\rho}$, which allows 
for better comparison with results obtained in previous literature, as there such a
relation is often used (as for example in \citet{2008MNRAS.385.1211P}).\\

\subsection{Synchrotron emission from secondary models}\label{theory}

We consider a spectral distribution of the 
population of relativistic CRp described 
by a power law in energy $E_{\mathrm{p}}$ :

\begin{eqnarray}
   N(E_{\mathrm{p}}) &=& 
   K_{\mathrm{p}}E_{\mathrm{p}}^{-\alpha_{\mathrm{p}}} \, ;
\end{eqnarray}
if not stated otherwise, we use a spectral index 
$\alpha_{\mathrm{p}} = 2.6$.
The normalisation $K_{\mathrm{p}}$ is model dependent and is
defined in the next Sections. \\ 
The main channel of hadronic interaction between the CR protons 
and the ambient medium is multi-pion production \citep{1999APh....12..169B}: 
\begin{eqnarray*}
	p_{\mathrm{CR}}+p_{\mathrm{th}} &\Rightarrow&
        \pi^{0}+\pi^{+}+\pi^{-} + \mathrm{anything}\\ \pi^{\pm}
        &\Rightarrow& \mu + \nu_{\mu}  \\ \mu^{\pm} &\Rightarrow&
        e^{\pm} + \nu_{\mu} + \nu_{e} \\  \pi^{0}	 &\Rightarrow&
        2\gamma\\
\end{eqnarray*}

\noindent
In this case the 
injection spectral rate of secondary $e^{\pm}$ is 
\citep[e.g.][]{2005MNRAS.363.1173B} :

\begin{align}
Q_{e}(E)&=
\int_{E_{\pi}}
Q_{\pi}(E_{\pi}) dE_{\pi} \times \nonumber \\
&\times \int dE_{\mu} F^{\pm}_{e}(E_{\pi},E_{\mu},E_e)
F_{\mu}(E_{\mu},E_{\pi}),
\label{qepm1}
\end{align}

\noindent
where $F_e^{\pm}(E_e,E_\mu,E_\pi)$ is the spectrum of
electrons and positrons from the
decay of a muon of energy $E_\mu$ produced in the decay of a pion with
energy $E_\pi$ (taken from \citet{1999APh....12..169B},
$F_{\mu}(E_{\mu},E_{\pi})$ is the muon spectrum
generated by the decay of a pion of energy $E_{\pi}$ 
\citep[e.g.][]{1998ApJ...493..694M}, and the
pion injection rate due to p-p collisions is
\citep[e.g.][]{1986A&A...157..223D,1999APh....12..169B,2005MNRAS.363.1173B}

\begin{align}
Q_{\pi}(E_{\pi}) &= n_{th} c 
\int dE_p N(E_p) \beta_p  F_{\pi}(E_{\pi},E_p) \, \times \nonumber \\
&\times\sigma_{pp}(E_p) \sqrt{1 - ({{m_p \, c^2}\over{E_p}})^2},
\label{q_pi}
\end{align}

\noindent
where $n_{th}$ is the number density of thermal protons,
$\sigma_{pp}$ is the p-p cross-section, 
and $F_{\pi}$ is the spectrum of pions from the collision between a CRp 
of energy $E_p$ and thermal protons 
\citep[e.g.][]{1998ApJ...493..694M,1999APh....12..169B,2005MNRAS.363.1173B}.

\noindent
The observed synchrotron emission in radio halos
at frequencies of $\approx \, \mathrm{GHz}$ requires energies of the
secondary particles of several GeV, considering field
strengths of $\approx \, \mu \mathrm{G}$ typical of the IGM
\citep[e.g.][]{2002ARA&A..40..319C}.
In this case a scaling model is appropriate to describe the pion-spectrum
from p-p collisions, and we calculate $Q_e(E)$ from 
Eqs.~\ref{qepm1}--\ref{q_pi} following \citet{2005MNRAS.363.1173B} and 
using a pion-spectrum :

\begin{equation}
F_{\pi}(E_{\pi})=
{1 \over {2 E_{\pi}}}
\left[ c_1 (1 - {{E_{\pi}}\over{E_p}})^{3.5} +
c_2 \exp (-18 {{E_{\pi}}\over{E_p}}) \right]
\label{fpi}
\end{equation}

\noindent where $c_1$=1.22 and $c_2$=0.92 \citep{1990ApJ...349..620B}.

\noindent
The resulting electron source function can be approximated by :

\begin{equation}
Q_{\mathrm{e}^{\pm}}(E) =
c \,  n_{\mathrm{th}} 
K_{\mathrm{p}} E^{-\alpha_{\mathrm{p}}}
f(E, \alpha_{\mathrm{p}})
\label{qelectrons}
\end{equation}

\noindent
where $f(E, \alpha_{\mathrm{p}})$ accounts for the log--scaling
of the p-p cross--section at high energies and causes the spectral shape
to be slightly flatter than $E^{- \alpha_{\mathrm{p}}}$;
an analytical expression for the asymptotic form of $f$ (for 
$E_{\mu} >> m_{\mu} c^2$, $E_{\pi} >> m_{\pi} c^2$,
$E_{p} >> m_{p} c^2$) is
derived in  \citet{2005MNRAS.363.1173B}.

The steady state spectrum of high energy 
electrons in the IGM is given by \citep[e.g.][]{2000A&A...362..151D}:

\begin{eqnarray}
	N_{\mathrm{e}}(E) &=& \left|
        \dot{E}(E) \right|^{-1}
        \int\limits_{E}^{\infty} dE' \,
        Q_{\mathrm{e}^{\pm}}( E')
\label{Nstat}
\end{eqnarray}

\noindent
where the relevant cooling 
processes involve synchrotron and inverse Compton losses:
\begin{eqnarray}
	\dot{E}(E) &=& - \frac{4
          \sigma_{\mathrm{T}}c}{3 m_{\mathrm{e}}^{2} c^{4}} \left(
        \frac{B^{2}}{8\pi} + \frac{B^{2}_{\mathrm{CMB}}}{8\pi} \right)
        E^{2},
\label{CRloss}
\end{eqnarray}
$B$ is the local magnetic field strength and
$B^{2}_{\mathrm{CMB}}/8\pi$ gives the energy density of the CMB expressed
as an equivalent magnetic field.

\noindent 
We calculate the electron spectrum from Eqs. \ref{qepm1}--\ref{fpi} and
\ref{Nstat}--\ref{CRloss}; this can be approximated by :

\begin{equation}
N_{\mathrm{e}}(E)= {{6 \pi m_e^2 c^4}\over{\sigma_T}}
n_{th} K_p E^{-(1+\alpha_{\mathrm{p}})} 
{{ g(E,\alpha_{\mathrm{p}}) }\over
{B^{2} + B^{2}_{\mathrm{CMB}}}}
\label{ne_stat}
\end{equation}

\noindent
where $g$ is related to $f$ in Eq. \ref{qelectrons} via :

\begin{align}
g(E,\alpha_{\mathrm{p}}) &=
{1 \over{ E^{1 - \alpha_{\mathrm{p}}} }}
\int_E dX X^{-\alpha_{\mathrm{p}}} 
f(X, \alpha_{\mathrm{p}}) \nonumber \\
&\propto E^{\Delta}
\end{align}

\noindent
and $\Delta \approx 0.2$ for $E \approx$ few GeV.

\noindent
The radio emissivity is :
\begin{align}\label{syn_emiss}
	j_{\nu} &= \frac{\sqrt{3}e^{3}B }{m_{\mathrm{e}}
          c^{2}}\int\limits^{E_{\mathrm{max}}}_{E_{\mathrm{min}}}
        \int\limits_{0}^{\frac{\pi}{2}}\mathrm{d}E_{\mathrm{e}}
        \mathrm{d}\theta  \, \sin^{2}\theta
        F\left(\frac{\nu}{\nu_{c}}\right)
        N_{\mathrm{e}}(E)
\end{align}
where $\nu_{\mathrm{c}} = (3/4\pi) p^{2} e B \sin \theta /(mc)^{3}$ is
the critical frequency and F the integral over the synchrotron kernel
:
\begin{align}
	F(x) &=	x\int\limits^{\infty}_{x}
        K_{\frac{5}{3}}(\xi)\,\mathrm{d}\xi .
\end{align}
Here $K_{\frac{5}{3}}$ denotes the modified Bessel function of order
5/3.

\noindent
For a power law spectrum of CRp, 
$N_p(E_p) \propto E_p^{-\alpha_{\mathrm{p}}}$, the resulting
synchrotron emission from secondary electrons
is $j_{\nu} \propto \nu^{-(\alpha_{\mathrm{p}} -\Delta)/2}$.

\subsection{$\gamma$-rays from hadronic interactions}\label{form_gamma}

As already mentioned before, CRp - proton collisions produce neutral
pions, which in turn decay to 2 photons. $\gamma$-rays
are a direct measure of the CRp and provide a 
complementary constraint to the injection process of secondary
electrons.

\noindent
In order to allow for a prompt comparison with recent results we  
follow the formalism described in
\citet{2004A&A...413...17P} to estimate the $\gamma$-ray flux from CR
in our simulations. 

The $\gamma$-ray source function is :
\begin{eqnarray}
q_{\gamma}(E_{\gamma})\, \approx {{ 2^{4-\alpha_{\gamma}} }\over{
3\alpha_{\gamma}}} 
{{\sigma_{\mathrm{pp}} c n_{\mathrm{th}} K_p }\over{ \alpha_{\mathrm{p}} -1}}
(E_{\mathrm{p,min}})^{-\alpha_{\mathrm{p}}} \frac{E_{\mathrm{p,min}}}{\mathrm{GeV}} \times
\nonumber\\
\left(
\frac{m_{\pi^{0}}c^{2}}{\mathrm{GeV}}
\right)^{\alpha_{\gamma}} \left[ \left(
\frac{2E_{\gamma}}{m_{\pi^{0}}c^{2}}\right)^{\delta_{\gamma}}
+ \left(\frac{2E_{\gamma}}{m_{\pi^{0}}c^{2}}\right)^{-\delta_{\gamma}}
\right]^{-\alpha_{\gamma}/\delta_{\gamma}}
\end{eqnarray}
where $\alpha_{\gamma} \simeq \alpha_{\mathrm{p}}$ is the asymptotic slope of 
the $\gamma$-ray
spectrum, which resembles the slope of the proton spectrum
\citep{1986A&A...157..223D}. The shape parameter, which describes the
semianalytic model near the pion threshold, is 
$\delta_{\gamma} = 0.14 \alpha_{\gamma}^{-1.6} + 0.44$ by using an
effective cross-section $\sigma_{\mathrm{PP}} = 32 \times (0.96 +
\mathrm{exp}(4.4-2.4\alpha_{\gamma})) \,\mathrm{mbarn}$. 
\noindent
The integrated $\gamma$-ray source density $\lambda_{\gamma}$ is then
obtained by integrating the source function
over energy \citep{2004A&A...413...17P} :

\begin{align}
	\lambda_{\gamma} &= \int\limits_{E_{1}}^{E_{2}}
        \mathrm{d}E_{\gamma} q_{\gamma}(E_{\gamma}) \nonumber \\ &=
        \frac{\sigma_{\mathrm{pp}} m_{\pi}c^{3}}{3
	\alpha_{\gamma}\delta_{\gamma}} 
	\frac{n_{\mathrm{th}} K_{\mathrm{p}}}{\alpha_{\mathrm{p}} -1} 
	\frac{ (E_{\mathrm{p,min}})^{-\alpha_{\mathrm{p}}}}{2^{\alpha_{\gamma}-1}}
        \frac{E_{\mathrm{p,min}}}{\mathrm{GeV}}  \left(
        \frac{m_{\pi_{0}}c^{2}}{\mathrm{GeV}}
        \right)^{-\alpha_{\gamma}} \nonumber \\ &\times
        \left[\mathcal{B}_{\mathrm{x}}\left(
          \frac{\alpha_{\gamma}+1}{2\delta_{\gamma}},
          \frac{\alpha_{\gamma}-1}{2 \delta_{\gamma}} \right)
          \right]^{x_{1}}_{x_{2}} 
\end{align}
where $\mathcal{B}_{\mathrm{x}}(a,b)$ denotes the incomplete
beta-function and $[f(x)]^{a}_{b} = f(a) - f(b)$.

\subsection{The three models}

To investigate the dependence of the predicted properties of non thermal
emission of clusters on the underlying assumptions, we investigate 3
models for the distribution of magnetic fields and cosmic rays in clusters.
They are chosen to encompass the reasonable range suggested by theoretical 
and observational findings.

\noindent
We keep the spectral index fixed to $\alpha_{\mathrm{p}}=2.6$ in order to 
be able to match the typical spectrum of giant radio halos, 
$\alpha \sim 1.2-1.3$ \citep[e.g.][]{2008SSRv..134...93F}, although 
a fraction of presently known halos has a steeper spectrum 
\citep[e.g.][]{2008Natur.455..944B,2009A&A...507.1257G}
Also, a spectral index $\alpha_{\mathrm{p}}=2.6$ allows to fit the 
spectral shape of the Coma halo at $\nu \leq 1.4$ GHz, although also in this
case the spectrum steepens at higher frequencies 
\citep[e.g.][]{2003A&A...397...53T,2010MNRAS.401...47D}

\subsubsection{Model 1: Constant $X_{\mathrm{CR}}$}\label{const_xcr}

\begin{figure*}
\includegraphics[width=0.9\textwidth]{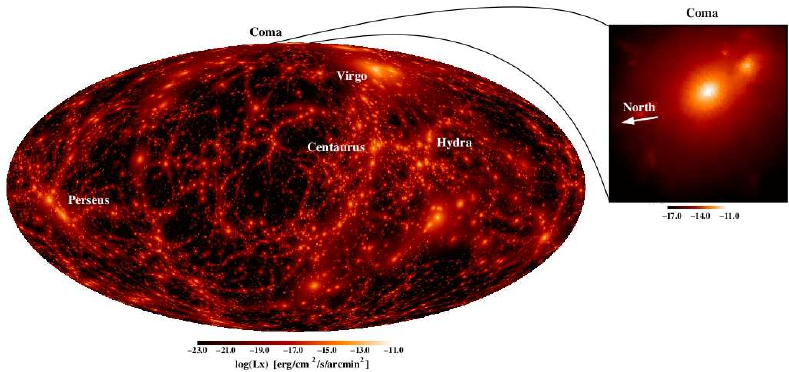}
\includegraphics[width=0.9\textwidth]{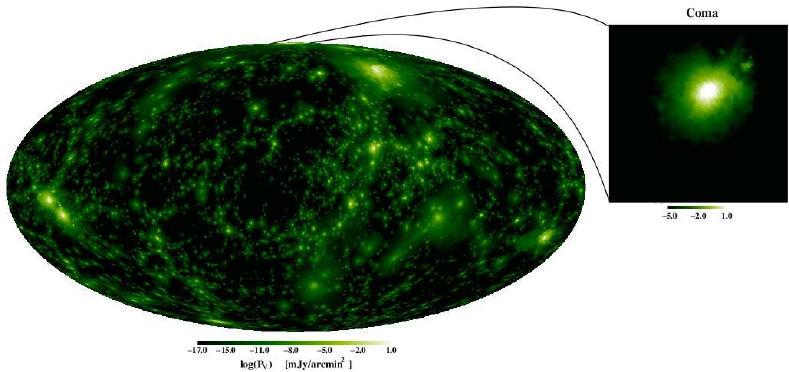}
\includegraphics[width=0.9\textwidth]{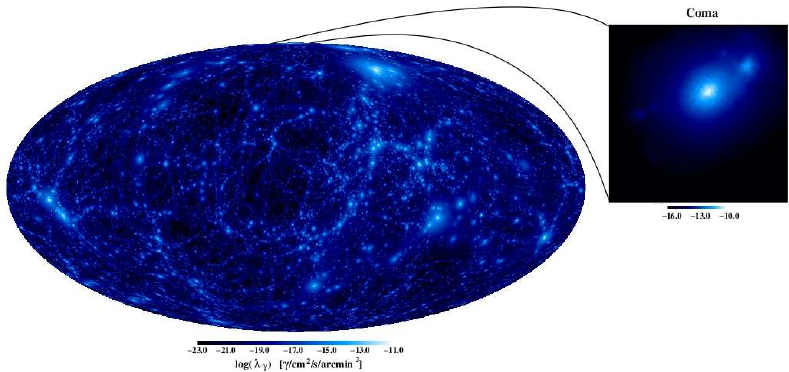}
\caption{Full sky maps of the simulation in galactic cordinates. 
From top to bottom, X-ray, radio and $\gamma$-ray surface brightness
are shown. For the radio and $\gamma$-ray emission the model with the 
constant cosmic ray energy fraction was chosen. The $\gamma$-ray emission 
is evaluated for VERITAS (e.g. $E>100\,GeV$).
The inlay shows a zoom onto a 3$\times$3 degree region around 
the Coma cluster. In the upper most map, the most prominent clusters 
of the local universe are labled and the arrow in the inlay points 
towards north.}\label{obs}
\end{figure*}

In our first model the energy density of the CR protons is taken 
as a constant fraction of the thermal energy density. 
This is reasonable if a constant fraction of the energy that is
channeled into the IGM to heat the gas goes into the acceleration of CRp.

Therefore, in this model, the normalisation $K_{\mathrm{p}}$ is chosen 
to have a constant fraction,  $X_{\mathrm{p}}=\mathrm{const}$, 
of kinetic CRp 
energy density $\epsilon_{\mathrm{p}}$ to thermal energy 
density $\epsilon_{\mathrm{th}}$ of the IGM: 
\begin{eqnarray}\label{crpscaling}
	\epsilon_{\mathrm{p}} &=& X_{\mathrm{p}}\,
        \epsilon_{\mathrm{th}} \\
		  &=& \frac{K_{\mathrm{p}}}{\alpha_{\mathrm{p}}-2}
		  \left( E_{\mathrm{p,min}} \right)^{2-\alpha_{\mathrm{p}}}
\end{eqnarray}

For the magnetic field distribution within the IGM we take directly 
the magnetic field extracted from the simulations.

\subsubsection{Model 2: Varying $X_{\mathrm{CR}}(r)$}\label{varxcr}

In a second model we adopt a radius dependent cosmic ray energy 
density fraction, $X_{\mathrm{CR}}(r)$, as obtained from simulations 
of CR acceleration in structure formation shocks by 
\citet{2007MNRAS.378..385P}. We also assume a constant CRp spectral 
index over the whole cluster volume, $\alpha_{\mathrm{p}}=2.6$.

\begin{figure}
   \centering \includegraphics[width=0.45\textwidth]{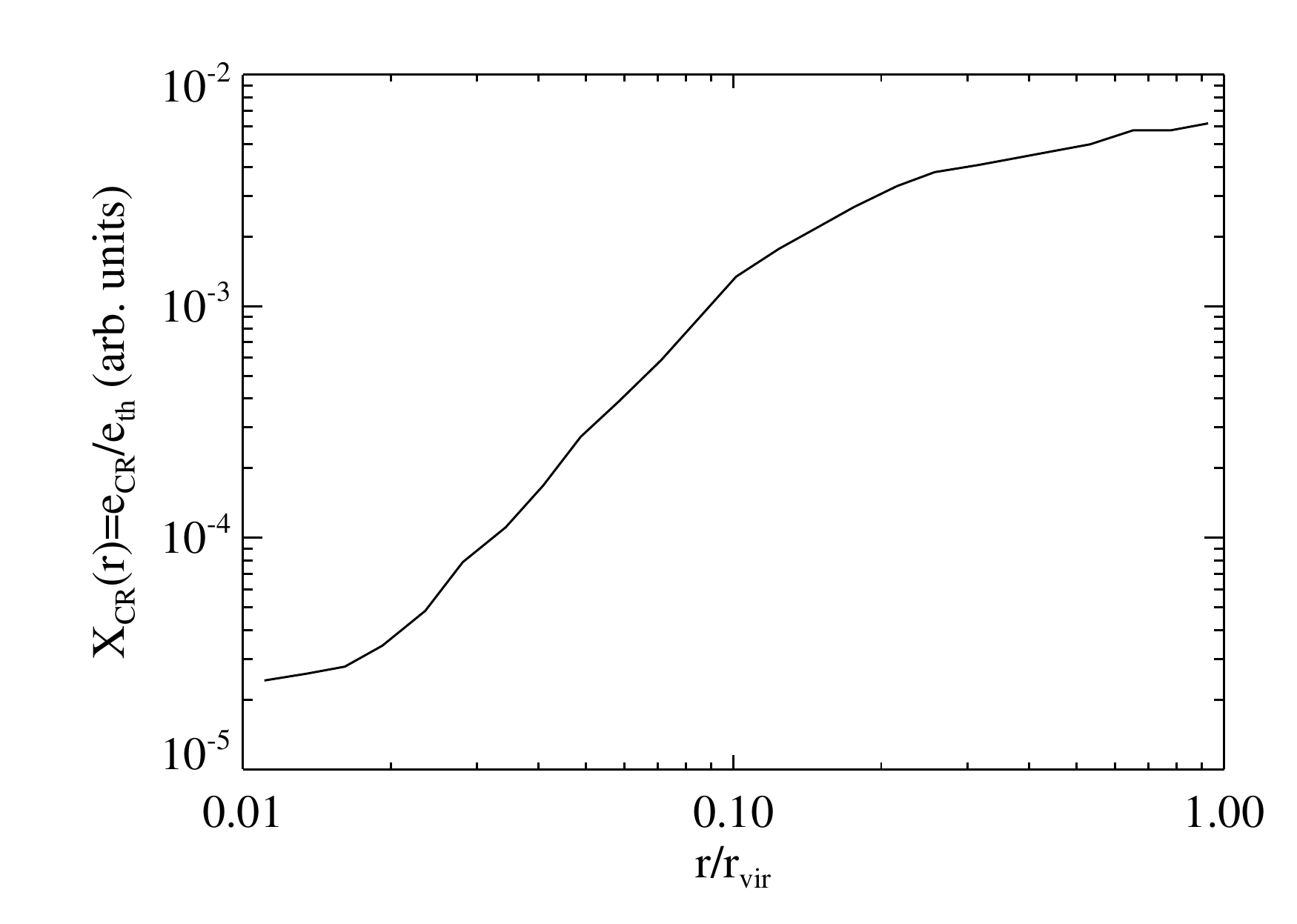}
	\caption{Energy in cosmic ray protons relative to thermal
          energy over cluster radius, inferred from
          \citet{2007MNRAS.378..385P}. The dependence was derived
          assuming a linear relation between pressure and energy
          density. }\label{img_xcr}
\end{figure}

\citet{2008MNRAS.385.1211P} simulated the injection of CR protons by
merger shocks during structure formation. They find that cosmic ray
pressure increases relative to thermal pressure with increasing
distance to the cluster centre. Assuming an ideal gas this directly
translates into a radially increasing cosmic ray energy density
fraction $X_{\mathrm{CR}}(r) = \epsilon_{\mathrm{th}}/
\epsilon_{\mathrm{p}}$. 
We use these results to infer a radially varying normalisation, so that
Eq.~\ref{crpscaling} becomes:
\begin{eqnarray}\label{crpscaling2}
\epsilon_{\mathrm{p}}({\bf r}) &=& X_{\mathrm{p}}(r)\,
\epsilon_{\mathrm{th}}({\bf r}) \\
&=& \frac{K_{\mathrm{p}}(r)}{\alpha_{\mathrm{p}}-2}
\left(E_{\mathrm{p,min}} \right)^{2-\alpha_{\mathrm{p}}}.
\end{eqnarray}

Figure \ref{img_xcr} shows the radial dependence of the fraction used
in agreement with the average over two cool core clusters, as shown in
\citet{2008MNRAS.385.1211P}. 
The fraction increases by more than 2 orders of magnitude in the
cluster external regions.
This is expected to increase the extension of the diffuse
radio emission generated by secondary electrons with respect to model 1.
For the magnetic field distribution within the IGM we again use
the magnetic field extracted from the simulations.

\subsubsection{Model 3: Scaling $B \propto \sqrt{\rho}$}

In our third model we modify the magnetic field originally 
obtained from the cosmological simulations.

\begin{figure}
\centering 
\includegraphics[width=0.45\textwidth]{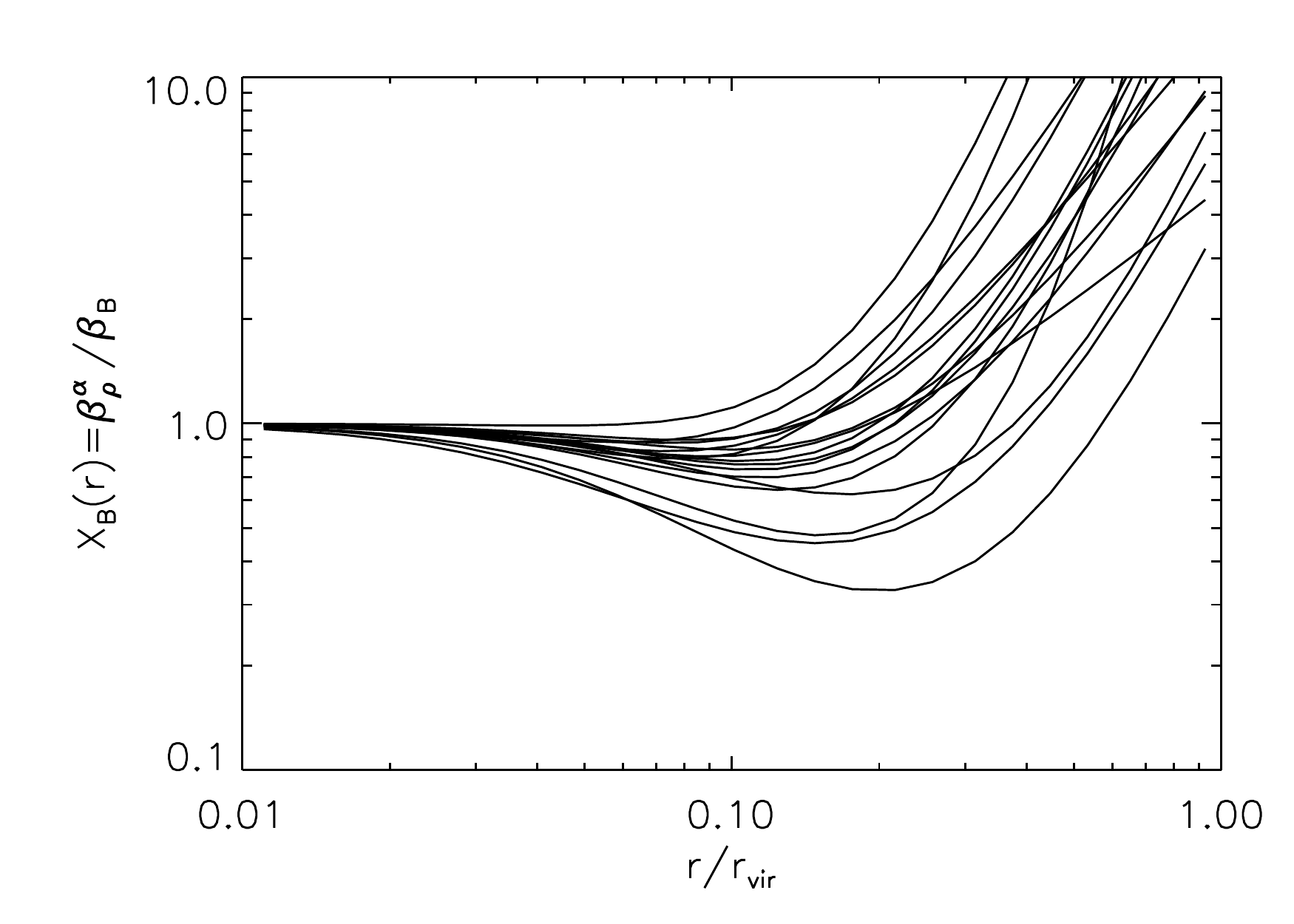}
\caption{Scaling functions $f_{\mathrm{scal}}(r)$ of all clusters. It
  modifies the magnetic field globally, so that its magnetic energy scales
  linear with thermal energy.}\label{bscal}
\end{figure}

The main reason for this is that the magnetic energy density,
$\epsilon_{\mathrm{B,0}}$, of the IGM is often modelled to follow 
thermal energy density, $\epsilon_{\mathrm{th}}$, 
which implies $B \propto \sqrt{\rho}$. 
In contrast to that we find $B \propto \rho$  in the outer regions of our simulated clusters. 
To allow a comparison with models that assume $B \propto \sqrt{\rho}$, 
we construct in our third model a function $f_{\mathrm{scal}}(r)$ which
scales the simulated magnetic field to follow $B \propto \sqrt{\rho}$ globally.
This allows us to alter the dependence of the field
with thermal density without loosing its inherent 
structure which is a result of our simulations. \par
To construct the scaling function, we fit beta models $n_{\beta} =
n_{0}(1+r^{2}/r_{\mathrm{core}}^{2})^{3\beta/2}$ to spherically
averaged density and magnetic field profiles. The scaling function is
then:
\begin{eqnarray}
	f_{\mathrm{scal}}(r) &=&
        \frac{\left(\rho_{0}(1+r^{2}/r_{\mathrm{core,\rho}}^{2})^{3/2\beta_{\rho}}\right)^{1/2}}{B_{0}(1+r^{2}/r_{\mathrm{core,B}}^{2})^{3/2\beta_{\mathrm{B}}}}
\end{eqnarray}
and is shown in figure \ref{bscal} for every cluster in the sample.
This changes the field strength at $r > 0.2 R_{vir}$ without strongly affecting
the field in the innermost regions.\\

\noindent
For the cosmic ray distribution in this model, $X_{\mathrm{p}}(r)$, we use the same radially dependent 
profile as in our model 2.

\section{Application to the cluster sample}\label{maps}

Having the diffuse synchrotron emission from secondary
electrons in our simulated clusters under control, 
the aim of this Section is to
compare the simulated properties with the most relevant
observational properties of radio halos.

\noindent
We use an unprecedented sample of fairly-massive simulated clusters made by 
the 16 most massive objects extracted from our simulations.
This allows for a statistical comparison with the main
observational properties of radio halos obtained from studies of volume 
limited samples of radio halos in galaxy clusters.
Additionally our sample (as well as the observational samples) 
contains the Coma cluster, so that we can calibrate $K_p$
in all our models to match the observed radio luminosity of the Coma 
halo where needed.

\subsection{The magnetic field in our cluster set}

The magnetic field obtained for our simulated Coma cluster is 
consistent with the one infered from modeling Faraday Rotation 
Measures of the Coma cluster, as shown
in \citep{2010MNRAS.401...47D}.

\begin{figure}
\centering 
\includegraphics[width=0.45\textwidth]{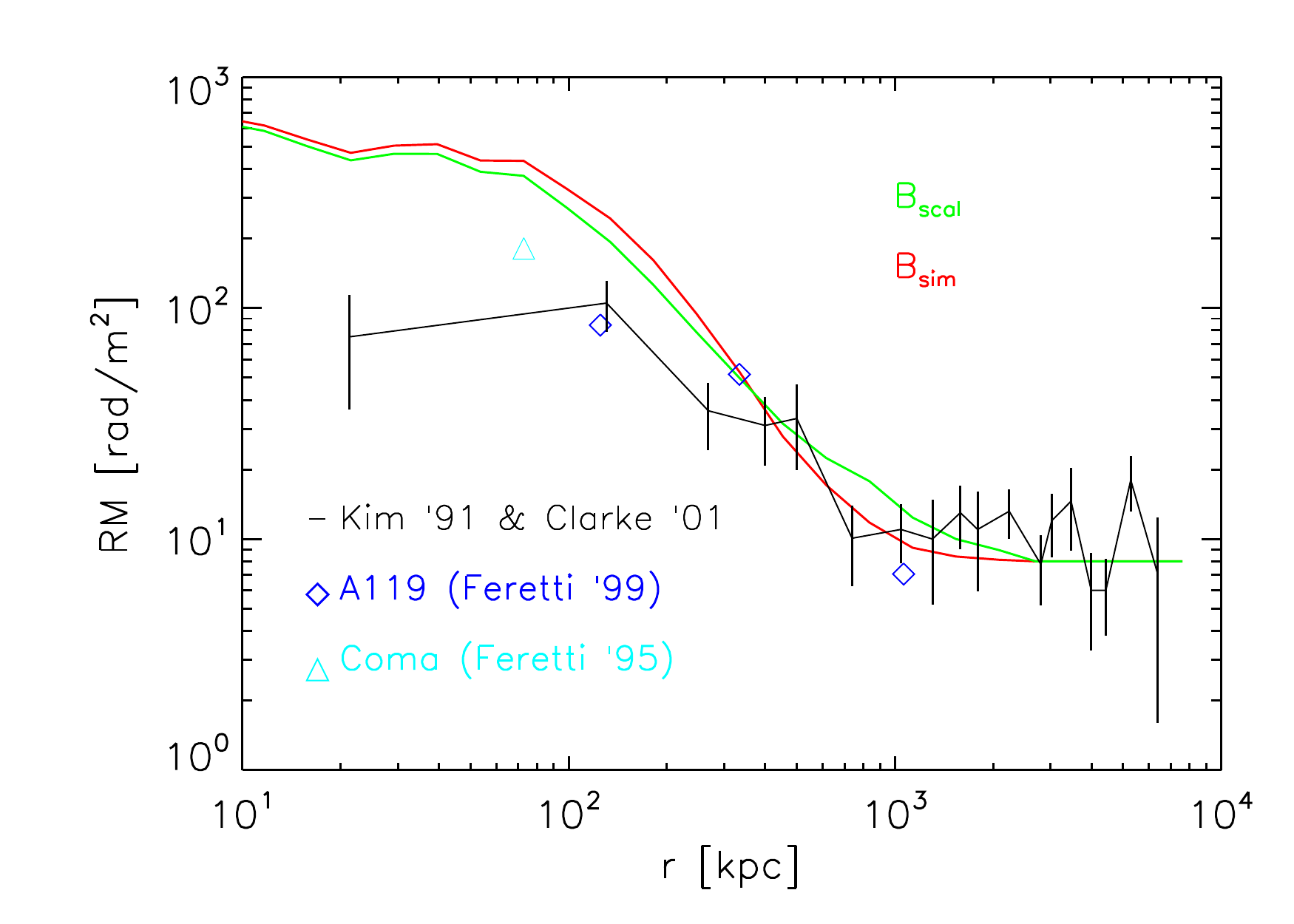}
\caption{Faraday rotation over radius for a mass selected
  subsample of our simulated clusters. Shown are the original
  (green) and upscaled (red) median over the whole sample in
  every radial bin. We also plot observations from Coma
  (turquoise), A119 (blue) and a sample of Abell clusters
  (black)
  \citep{1991ApJ...379...80K,2001ApJ...547L.111C,1995A&A...302..680F,1999A&A...344..472F}. The
  errors in Abell cluster sample were estimated using a
  bootstrapping technique (see text).}\label{bscal4}
\end{figure}

\noindent
To compare the ``statistical'' properties of the magnetic field 
in the complete set of simulated clusters with observations
and to show the effect of re-scaling the field (model 3), we
plot in figure \ref{bscal4} radial profiles
of Faraday Rotation.
Here we bin all clusters with gas mass $M > 3
\times 10^{14} \, M_{\odot}$ in radius and take the median in each
bin. In green we plot the field profile from simulations, and
in red the scaled field profile (model 3). 
To compare we overplot observations of a sample of Abell clusters 
from \citet{1991ApJ...379...80K,2001ApJ...547L.111C}, observations of
A119 from \citet{1999A&A...344..472F} and observations of the Coma
cluster \citep{1995A&A...302..680F}. 
Because of the small number of points,
the error from the Abell clusters was estimated using a bootstrapping
technique. For every bin we generate several samples, by selecting
every data point a random number of times and computing the median of
this subsample. We then compute the standard deviation of all samples
to get an estimate of the error in each bin.

\subsection{Radial profile of the radio emission}\label{rradprof}
 
\begin{figure*}
\centering
\includegraphics[width=0.33\textwidth]{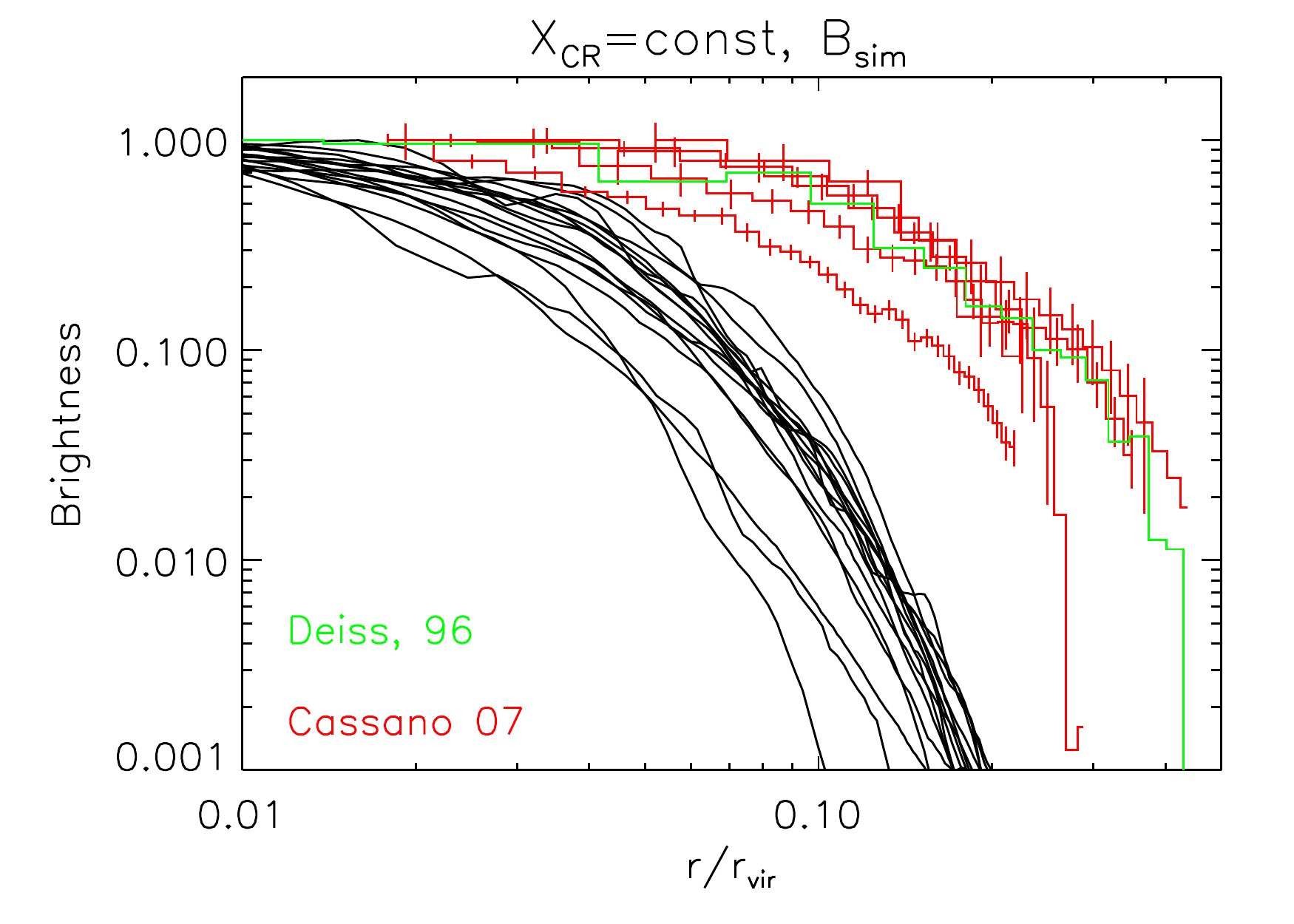}
\includegraphics[width=0.33\textwidth]{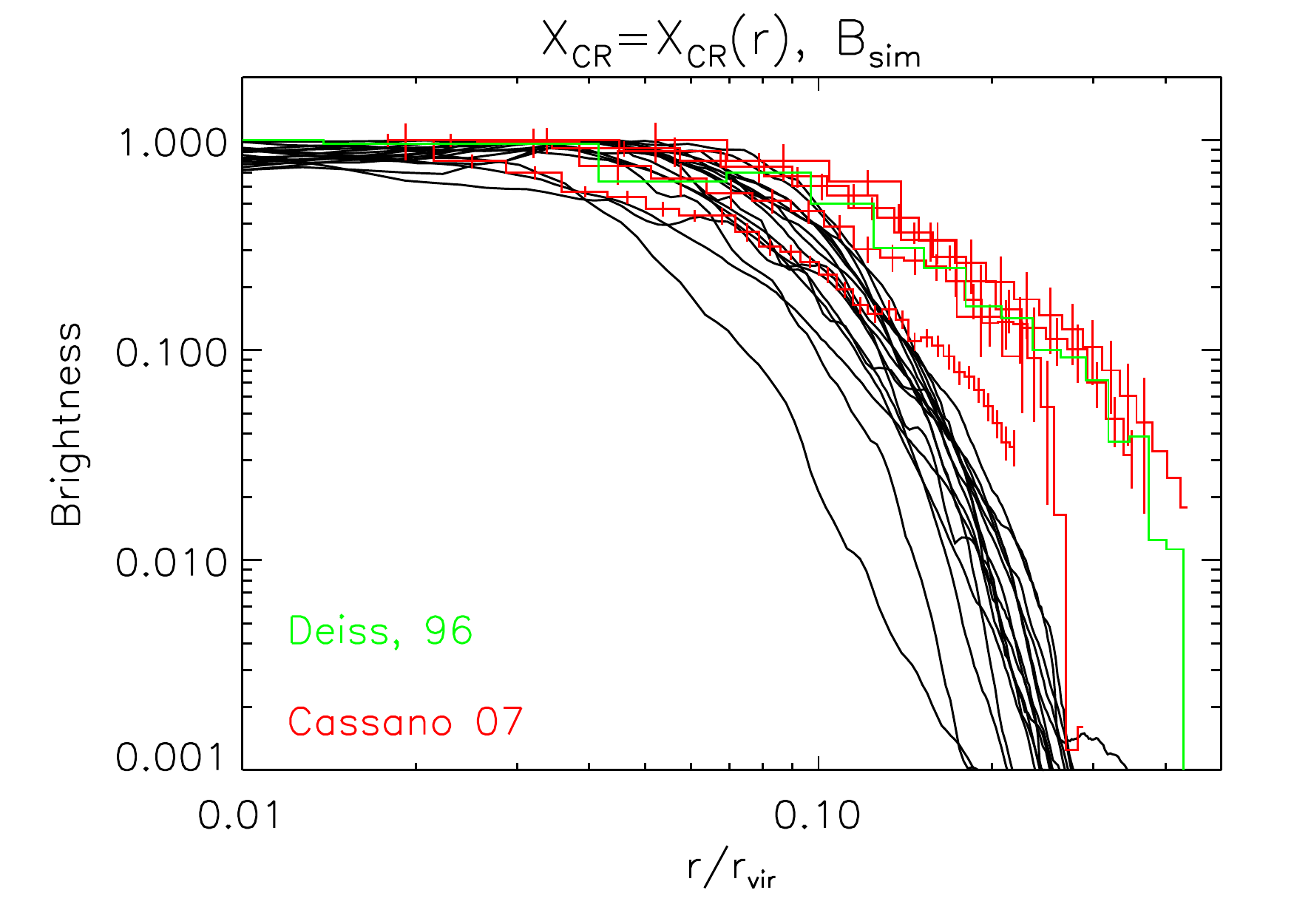}
\includegraphics[width=0.33\textwidth]{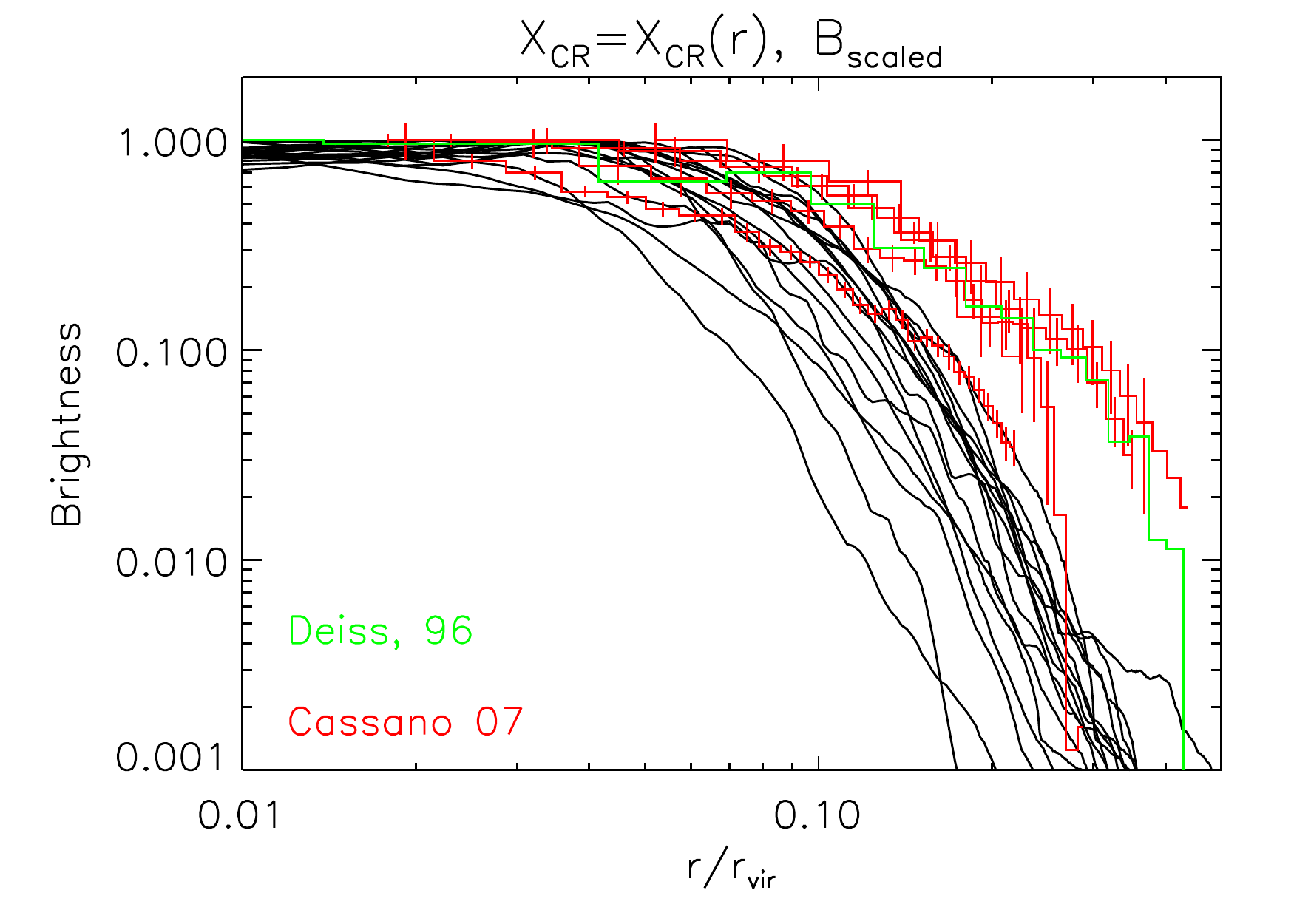}
\caption{Normalized radial profiles of radio emission from 17 simulated
  clusters (black). In the left panel the CRe population was
  modelled using a constant $X_{\mathrm{P}}$. The middle one
  uses the radius dependent CR to thermal energy fraction (fig. \ref{img_xcr})
  adopted from  \citet{2008MNRAS.385.1211P}. Additionally we modified (fig. \ref{bscal}) the
  magnetic field in the right panel to be
  $\propto \sqrt{\rho}$ (right panel). We ad observations  of
  A2744, A2319, A545, A2163 and A2255 from
  \citep{2007MNRAS.378.1565C} as red curves with error
  bars, Coma (green) is taken from
  \citet{1997A&A...321...55D}.}\label{radio_profiles}
\end{figure*}

Since both the target thermal-protons and the expected 
magnetic field strength in the IGM decrease with distance from
the cluster centre, most of the synchrotron luminosity emitted  
by secondary electrons should be produced in the cluster-core region.
This causes the radial profile of secondary-generated radio halos 
to be substantially steeper than those of the observed halos 
\citep[e.g.][]{2004MNRAS.350.1174B}, 
although formally this discrepancy may be alleviated 
by assuming that both the magnetic field and CRp have flat spatial 
distributions \citep{2004A&A...413...17P}.

\noindent
\citet{2007MNRAS.378.1565C} report radial profiles of the radio
emission of 5 well studied radio halos. 
We convolve our synthetic radio maps with a Gaussian using the typical 
beam size ($ 43\, \mathrm{kpc}$) of the observations, and in figure 
\ref{radio_profiles} compare these profiles with the observed ones.
Both simulated and observed profiles were normalised to one. 
We also include the radial profile of the Coma halo 
from \citet{1997A&A...321...55D}. 

\noindent
In all cases (models 1--3), 
Figure \ref{radio_profiles} shows that the radial profiles
of the secondary-generated radio halos in our simulated clusters are
considerably steeper than the observed ones.

\par In agreement with \citet{2004MNRAS.350.1174B}, by 
assuming a constant $X_{\mathrm{P}}$ (first panel), the simulated radio 
emission (black) for $r \ge 0.15 R_{\mathrm{vir}}$ is about 100 times 
below that measured in real radio halos (red, green).

By allowing $X_{\mathrm{P}}$ to increase with radius as in
\ref{varxcr} results in an increase of the simulated emission for 
larger radii (figure \ref{radio_profiles}, middle panel), but still 
expectations account for $< 10 \%$ of the observed emission 
in real radio halos at $r \geq 0.15 R_{\mathrm{vir}}$.
Additional up-scaling of the magnetic field (model 3) further increases the
size of the simulated halos, especially at very large distances from the
cluster center, where our MHD simulations would predict a steeper scaling
of the magnetic field with gas density.
However this up-scaling only mildly increases the level of the 
expected emission at intermediate distances, $r \approx 0.1-0.3
R_{\mathrm{vir}}$, where the radio brightness of radio halos is constrained
by present observations.

In a recent paper \citet{2008MNRAS.385.1211P} claim that, according to cosmological 
simulations, the combined synchrotron emission from secondary electrons 
and primary electrons accelerated at large scale shocks may produce diffuse 
emission with a fairly broad spatial distribution.
Based on our results, the contribution from primary electrons (even
assuming \citet{2008MNRAS.385.1211P} results) is not expected to solve the discrepancy
between models and observations in Fig.~\ref{radio_profiles}.
Indeed, our model 3 is thought to mimic the secondary-generated
emission in \citet{2008MNRAS.385.1211P} and, based on their Figure 9, the 
contributions from primary electrons
at $r \approx 0.2 R_{\mathrm{vir}}$ 
is (at best) comparable with that of the secondary 
electrons leaving expectations well below observations\footnote{it is more 
difficult to evaluate the
ratio of secondary to primary electrons at 
$r \approx 0.3 R_{\mathrm{vir}}$ from Figure 9 in 
Pfrommer et al. due to the presence of a very bright shock-like
spot southest of the center of their simulated
cluster (see their Figures 7 and 8) that affects the azimuthal brightness
profile, but that is not indicative of an upturn in the distribution 
of the truly diffuse halo-emission}.

In principle, for each cluster, it would be possible to allow
the energy density of CRp to further increase at large distances from the
cluster center and find an {\it ad hoc} spatial distribution of CRp
that allows for matching the brightness profiles of radio halos.
However this would imply the untenable scenario in which CRp store very
large energy budget outside the cluster core, for example in the case of 
the Coma halo the energy density of CRp at 
$r \approx 0.2-0.3 R_{\mathrm{vir}}$ should be comparable to that of the 
thermal ICM \citep{2010MNRAS.401...47D}.

\subsection{Morphology of the radio emission}\label{morphcoma}
\begin{table}
    \centering
    \begin{tabular}{c|c|c|c} \hline
        Cluster & M1 & M2 & M3 \\ \hline
		  0 & 1.09 & 1.04 & 0.87 \\
		  PERSEUS & 1.33 & 1.10 & 0.90 \\
		  COMA & 1.54 & 1.18  & 0.98 \\
		  4 & 1.21 & 1.01 & 0.94 \\
	 \hline
    \end{tabular}
\caption{Slopes of the Radio vers. X-ray surface brightness correlation from patches for our three models and the four largest clusters. \citet{2001A&A...369..441G,2001A&A...373..106F} find slopes of Coma \& A2163: b=0.64, A2391: b=0.98 and, A2255: b=0.82 } \label{P2Pslope_table}
\end{table}

\begin{figure*}
\centering
\includegraphics[width=0.33\textwidth]{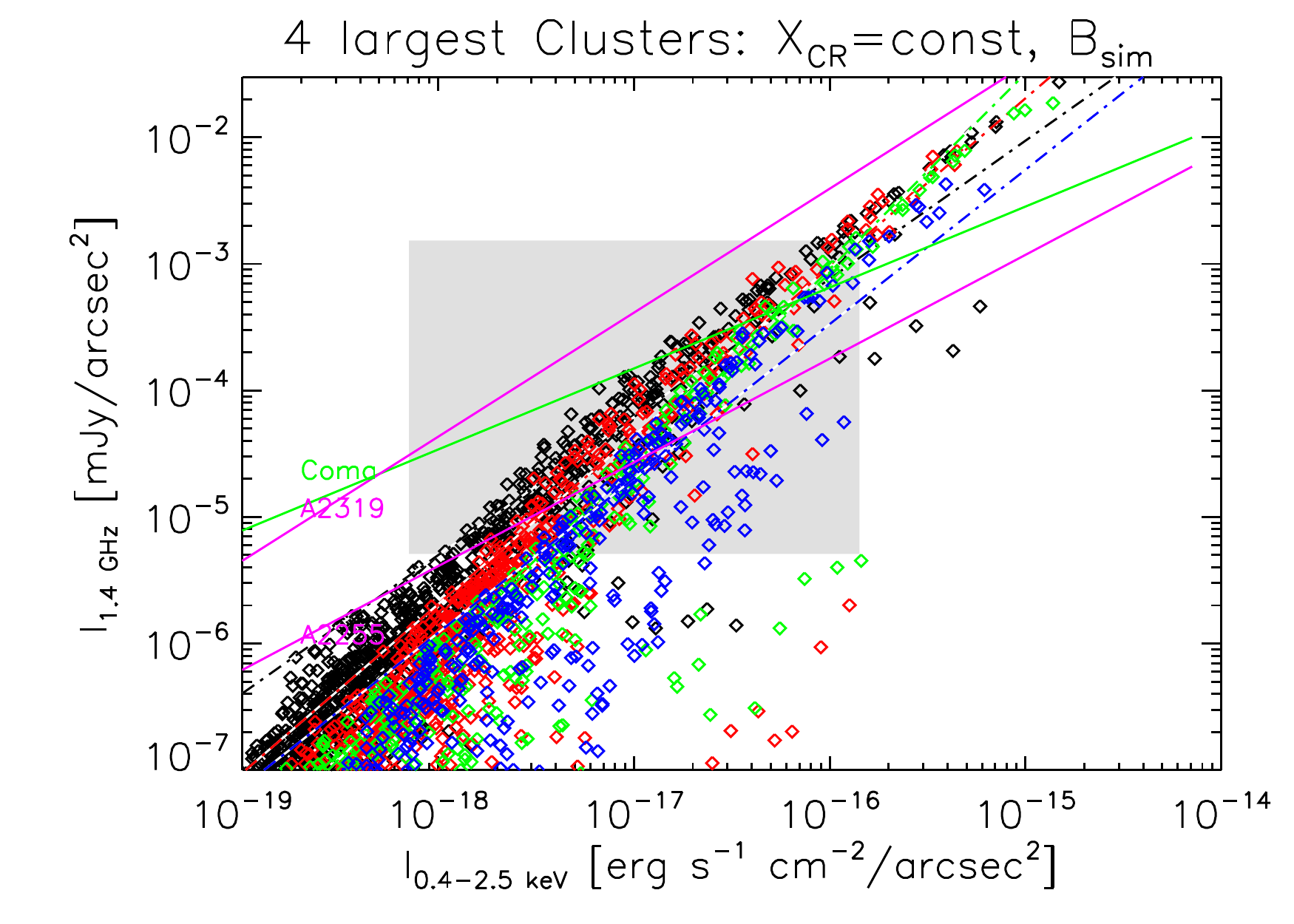}
\includegraphics[width=0.33\textwidth]{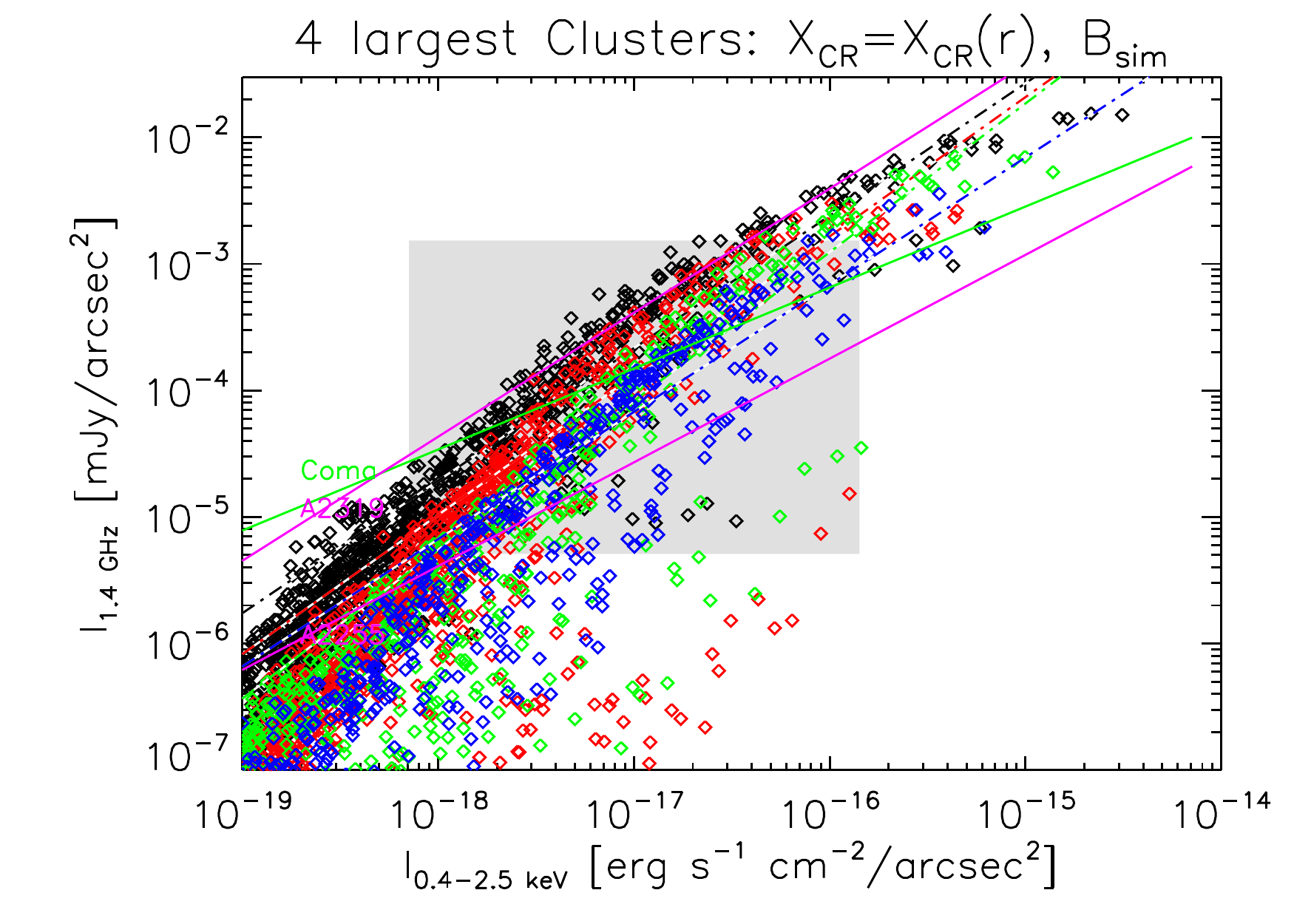}
\includegraphics[width=0.33\textwidth]{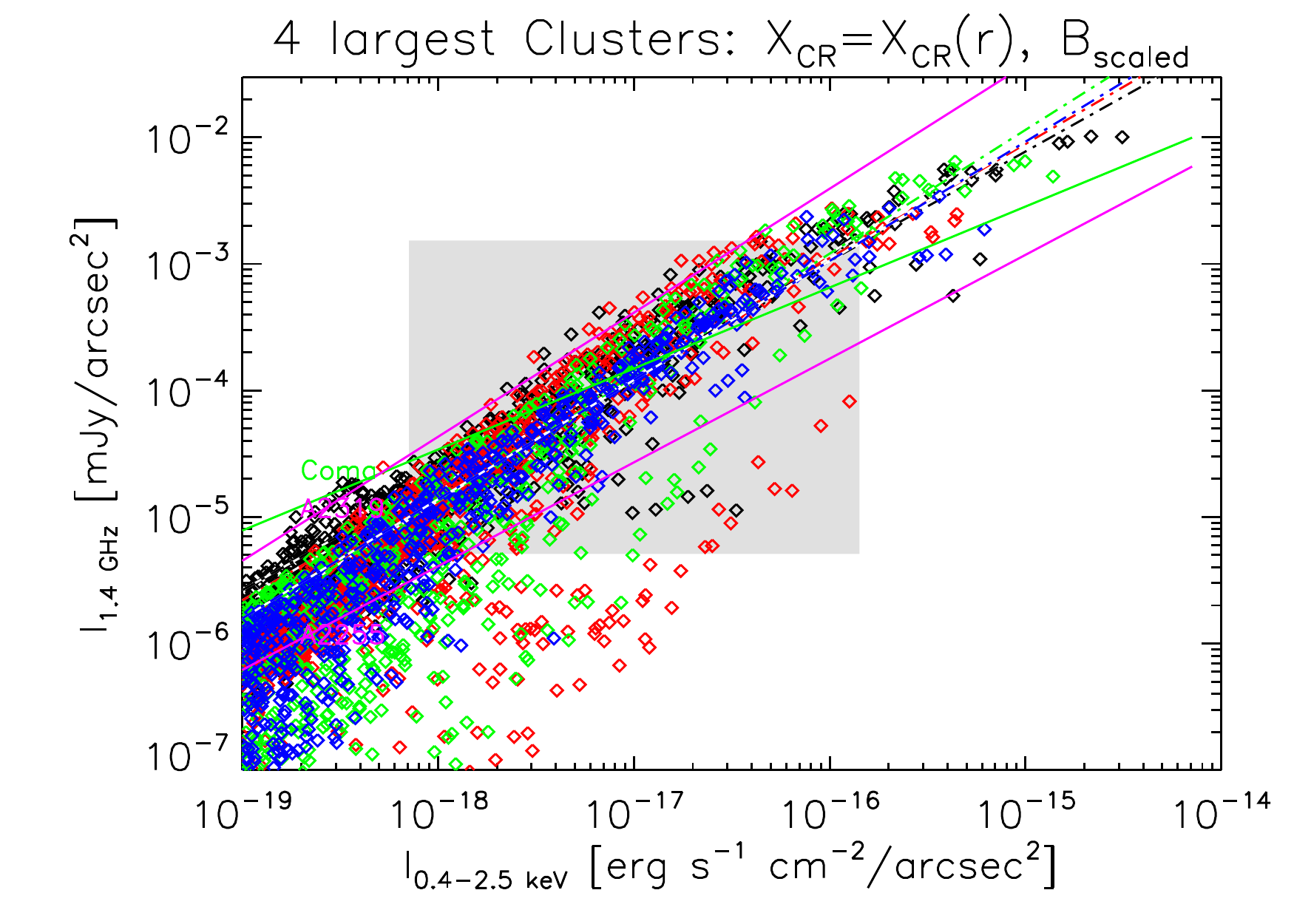}
\caption{Radio vers. X-ray surface brightness in patches with $166\, \mathrm{kpc}$
  side length, with the radio flux normalised to Coma.  We also plot a
  fit to observations from \citet{2001A&A...369..441G} (red, $b=0.64$
  for COMA and A2255,A2319 dotted).
The grey box indicates the observed region from
\citet{2001A&A...369..441G}.}\label{img_comamorph}
\end{figure*}

A complementary approach to compare expectations from 
simulations with observations is to  derive a point-to-point plot
between radio and X-ray brightness.
A number of clusters indeed show morphological correlations between X-ray and 
radio surface brightness \citep{2001A&A...369..441G}.
This could be preferable to a comparison of the
radial profiles because morphological distortions and imperfect
alignment of the emission centers may introduce errors in the
observed profiles \citep{2001A&A...369..441G}. 

\noindent
In Figure \ref{img_comamorph} we show 
the luminosity in $166\,\mathrm{kpc}$ sized
patches of our synthetic radio maps of the four biggest 
simulated clusters in our sample
as a function of the same patches in the X-rays. 
The three panels again show constant (model 1, left), 
increased cosmic ray scaling (model 2, middle) and cosmic rays scaled 
combined with magnetic fields scaled (model 3, right).  
We include the observed scaling for three
clusters (A2255, Coma, A2319) from \citet{2001A&A...369..441G} (dashed lines) and add the correlations found from best fits to the simulated data (dashed dotted lines).
In table \ref{P2Pslope_table} we include the slopes from a best fit model to the simulations inside the observed region. A comparison with the values found by  \citet{2001A&A...369..441G,2001A&A...373..106F}, (Coma, A2163: b=0.64, A2391: b=0.98, A2255: b=0.82) yields that only the third model may fit two of the three observations. Still no model approaches the flat value found for the Coma cluster.

\noindent
We find that none of the models fit the slopes of 
any of the observed scalings over
the whole range, with the radio brightness decreasing too quickly with
respect to the X-ray brightness. 
This reflects the point raised in the previous Section, i.e. that 
the slope of the radial distribution of the synchrotron emission
in our secondary-generated radio halos is too steep.

\noindent
In particular, 
the adoption of the cosmic ray scaling
function (models 2 and 3) causes the emission to decrease in the innermost 
five patches,
which can be seen in the plots as a bending at high brightness. 
The magnetic field scaling (model 3) adds more power to the outmost regions,
resulting in a flattening of the correlation, but still the obtained slope
is steeper than found in observations. \\  

\subsection{Scaling Relations}\label{scalrel}

There are several observed correlations for radio halos that relate 
thermal and non-thermal properties of the IGM: those between the 
radio power at 1.4 GHz, $P_{1.4}$, and the X-ray luminosity, $L_X$, 
temperature, and cluster 
mass \citep{2000ApJ...544..686L,2001A&A...369..441G,2006MNRAS.369.1577C}. 
In addition, by making use of a sample of 14 giant radio halos, 
\citet{2007MNRAS.378.1565C} found new scaling relations 
that connect the radio power, $P_{\mathrm{R}}$, of halos to the size 
of the emitting region, $R_{\mathrm{H}}$ 
\citep[see also][]{2009A&A...499..679M}, and to the total cluster 
mass within $R_{\mathrm{H}}$, $M_{\mathrm{H}}$;
a geometrical scaling was also found between $M_{\mathrm{H}}$ and
$R_{\mathrm{H}}$. 
The observed scalings from \citet{2007MNRAS.378.1565C} are :

\begin{align}
	P_{\mathrm{R}} &\propto R_{\mathrm{H}}^{4.18\pm0.68} \\ 
	P_{\mathrm{R}} &\propto M_{\mathrm{H}}^{1.99\pm0.22}\\ 
	M_{\mathrm{H}} &\propto R_{\mathrm{H}}^{2.17 \pm 0.19}
\end{align}

\noindent 
Specifically, $M_{\mathrm{H}}$ was computed from X-ray observations under the
assumption of hydrostatic equilibrium and spherical symmetry. 
This procedure may lead to errors as large as 40\% in mass
\citep{2006MNRAS.369.2013R} which are expected to be not dependent on
cluster mass, so that these errors might introduce considerable
scatter without affecting the real trend of the correlation.
$R_{\mathrm{H}}$ was measured on the radio images,
$R_{\mathrm{H}}=\sqrt{R_{\mathrm{min}}\times R_{\mathrm{max}}}$, 
where $R_{\mathrm{min}}$ and $R_{\mathrm{max}}$ are the minimum and
maximum radii measured on the 3$\sigma$ radio isophotes.
We stress that $R_{\mathrm{H}}$ provides a simple, but viable estimate of 
the physical size of radio halos, indeed a one-to-one correlation has 
been found between
$R_{\mathrm{H}}$ and the size containing the
85\% of the radio halo flux, $R_{85}$, derived from the observed
brightness profiles of halos \citep{2007MNRAS.378.1565C}.

A scaling was also found between the size of radio halos and the
virial radius of clusters, 
$R_{\mathrm{H}}\propto R_{\mathrm{vir}}^{2.63\pm0.50}$ \citep{2007MNRAS.378.1565C}. 
Given that massive clusters are almost self similar \citep[e.g.][]{2002ARA&A..40..539R} 
one might have expected that $R_{\mathrm{H}}$ scales with 
$R_{\mathrm{vir}}$ and that the radial
profiles of the radio emission are self-similar. On the contrary, this
result proves that self-similarity is broken in the case of the
non-thermal cluster components, as first noted by 
\citep{2001ApJ...548..639K}. 

\noindent
As the synchrotron power depends on both magnetic field scaling 
and CRe scaling with density, it is unclear what is responsible for 
the break in the observed properties.
On the other hand we know from previous work \citep{2008arXiv0808.0919D} that
the magnetic field scaling (with temperature or mass) flattens out for the 
largest clusters in our simulation, that would imply an expected break of
self similarity in the thermal vs non-thermal properties of our simulated
clusters.

\citet{2007MNRAS.378.1565C} showed that all the correlations explored 
so far for radio halos
can be derived by combining the $R_H-R_v$ and $P_{1.4}-R_H$ scalings. 
This suggests that there are two main scaling relations that 
carry out the leading information
on the physics of the non-thermal components in galaxy clusters.

In what follows we shall investigate whether the properties
of our simulated secondary-radio halos are consistent 
with the observed scalings.

\subsubsection{Mass vs. Size}
\begin{figure*}
\centering
\includegraphics[width=0.33\textwidth]{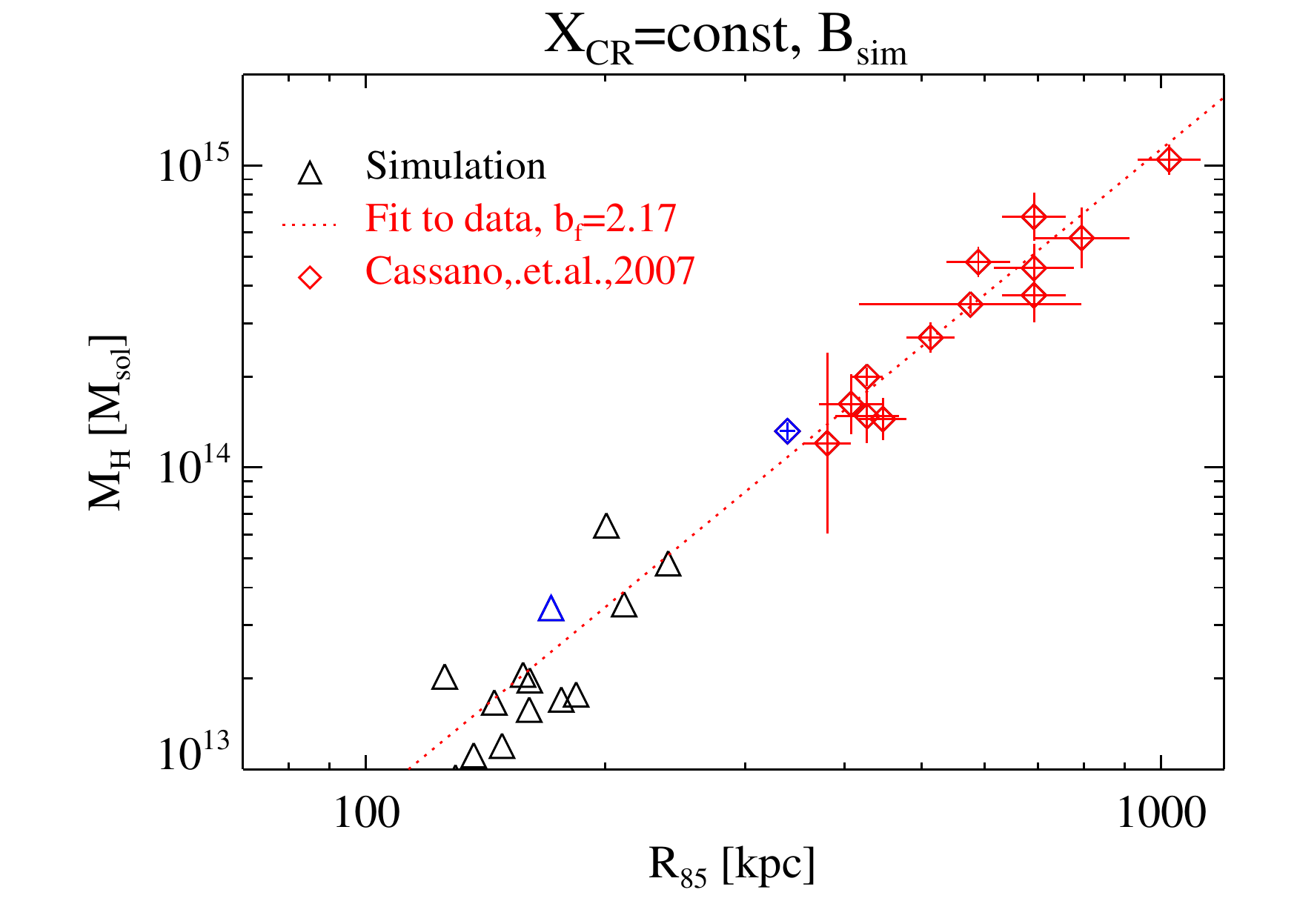}
\includegraphics[width=0.33\textwidth]{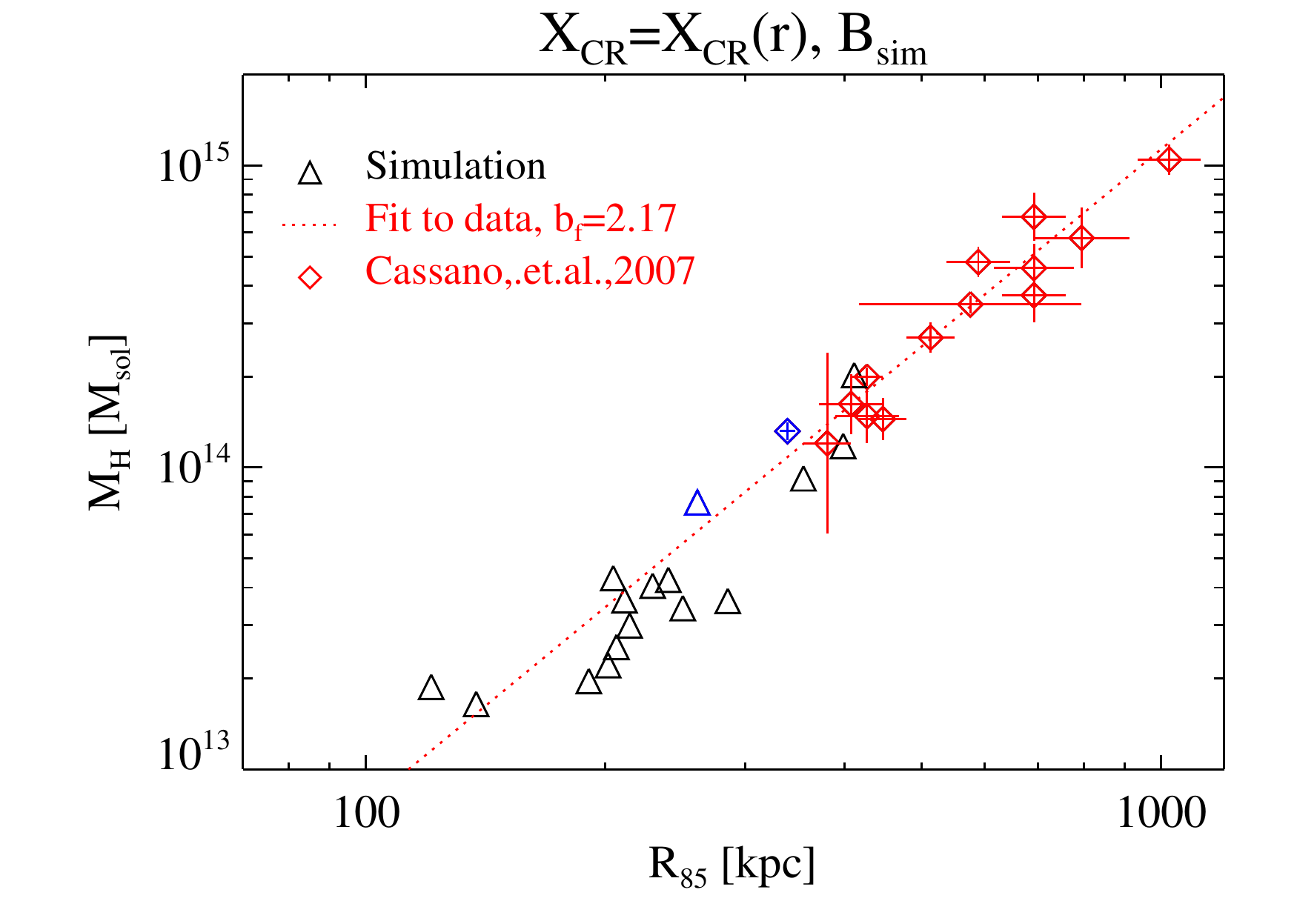}
\includegraphics[width=0.33\textwidth]{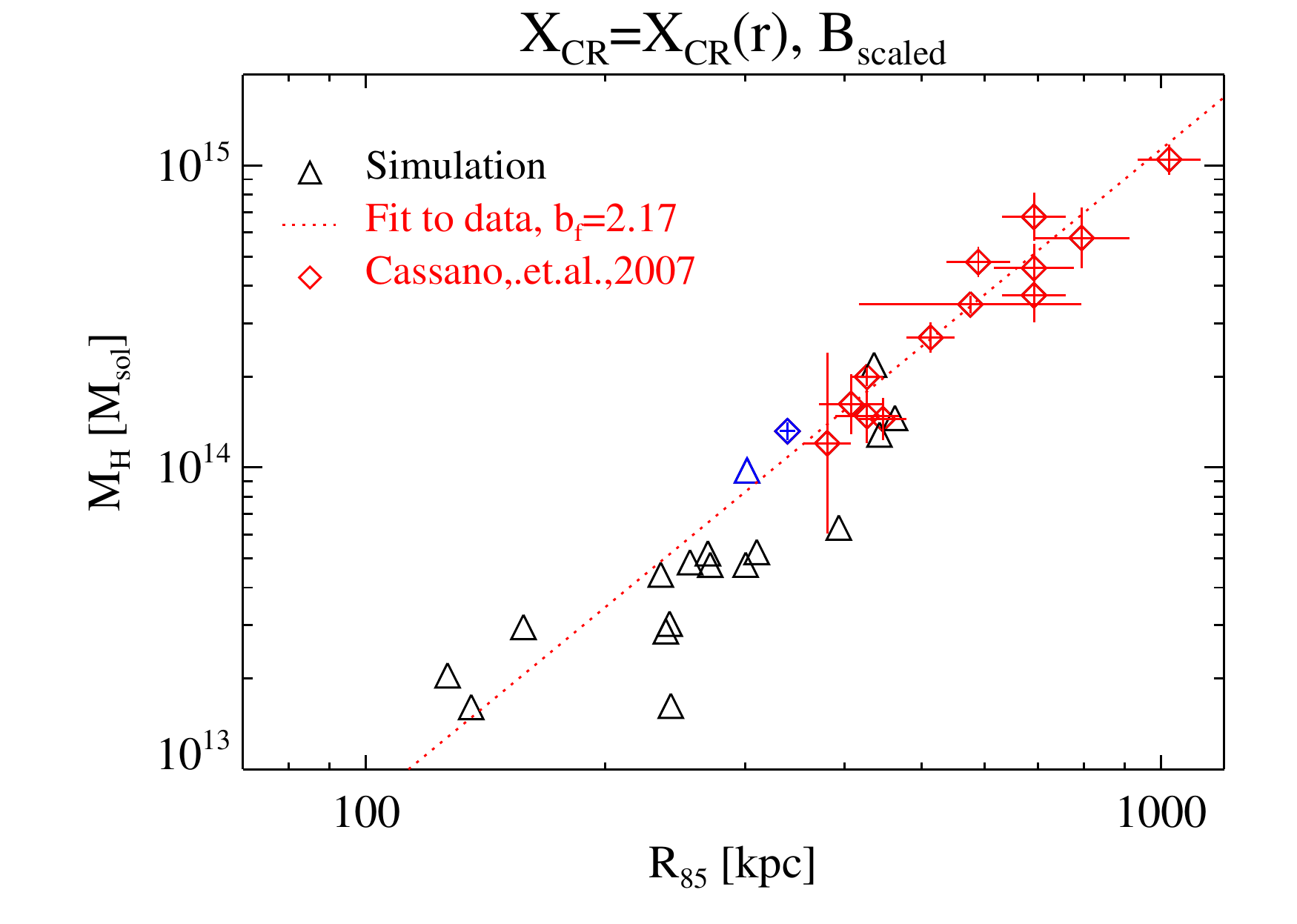}
\caption{Total gravitational mass inside the radio emitting region
  over radius of the same region. We plot the correlation for all
  three models, constant and varying CR fraction and varying CR
  fraction and upscaled magnetic field (from left to right). We also
  plot observations of 14 galaxy clusters from
  \citep{2007MNRAS.378.1565C} and the best fit correlations. Due to
  uncertainties in the mass estimation from X-ray data the
  observations may show systematic errors. For both, 
simulations and observations the coma cluster is marked in blue.
}\label{thermal_scal}
\end{figure*}

As a first step, 
before comparing the observed scalings with 
those derived for our simulated clusters, we check whether 
our clusters inherit the same mass distribution of real clusters 
with radio halos. 
We compare the observed and simulated scaling between the 
total mass inside $R_{\mathrm{H}}$ ($M_{85}$) and $R_{85}$, that provides
a geometrical scaling on the halo-region. 
Therefore we plot in Figure \ref{thermal_scal}
$M_{\mathrm{H}}$ versus $R_{85}$ for simulated (open triangles) 
and observed (diamonds) clusters, 
together with the best fit power law to the observed scaling 
from \citet{2007MNRAS.378.1565C}. 

The three models define different values of $R_{85}$ and consequently
different volumes where the scaling can be tested. 
We find that in all cases simulated clusters lie on the thermal 
scaling described by observed clusters, although, as expected 
(Sects.\ref{rradprof}--\ref{morphcoma}) simulations populate a region in the 
$M_{\mathrm{H}}$--$R_{85}$ diagram with smaller values of $R_{85}$.

\subsubsection{The Size vs. Size relation}
\begin{figure*}
\centering
\includegraphics[width=0.33\textwidth]{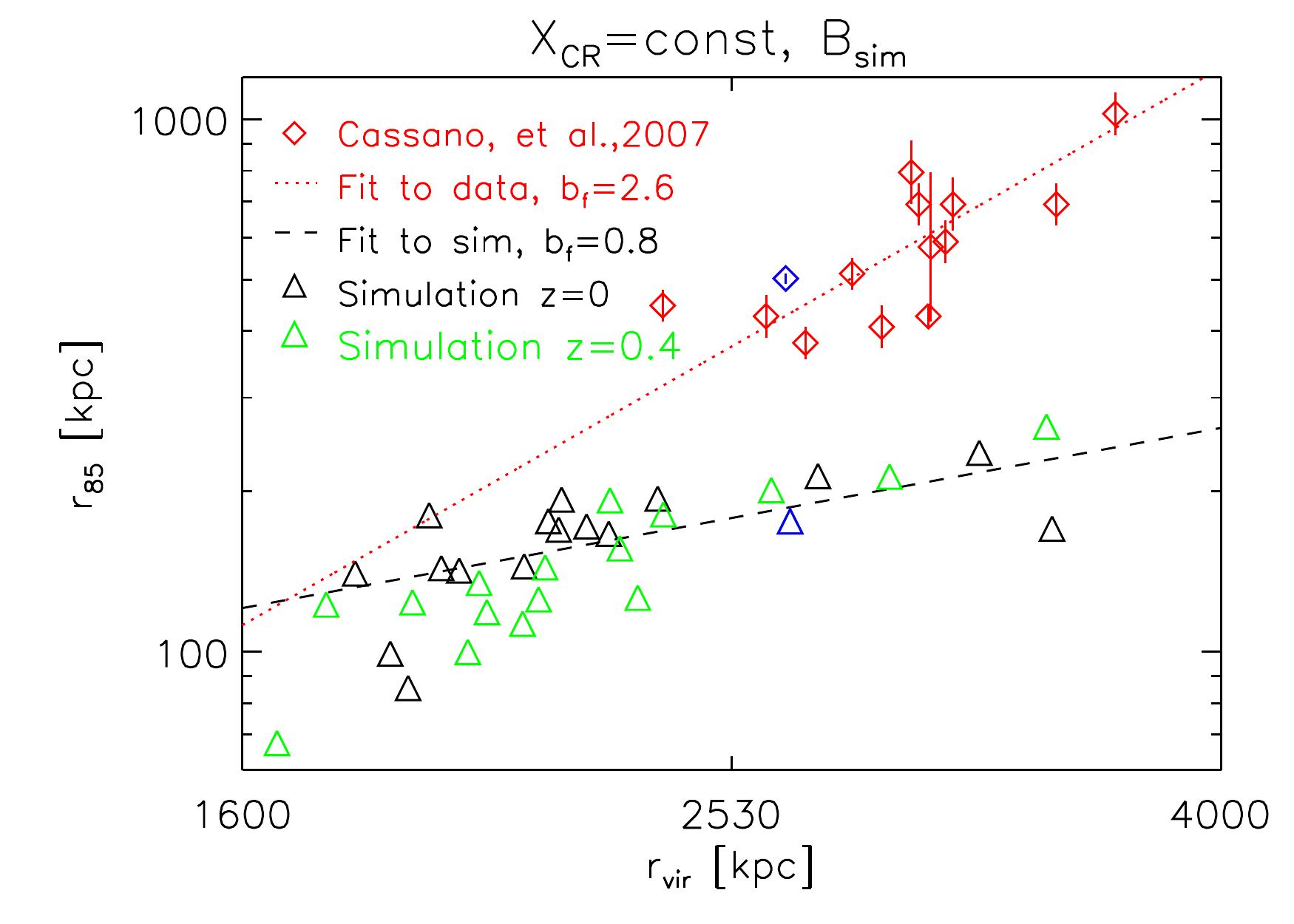}
\includegraphics[width=0.33\textwidth]{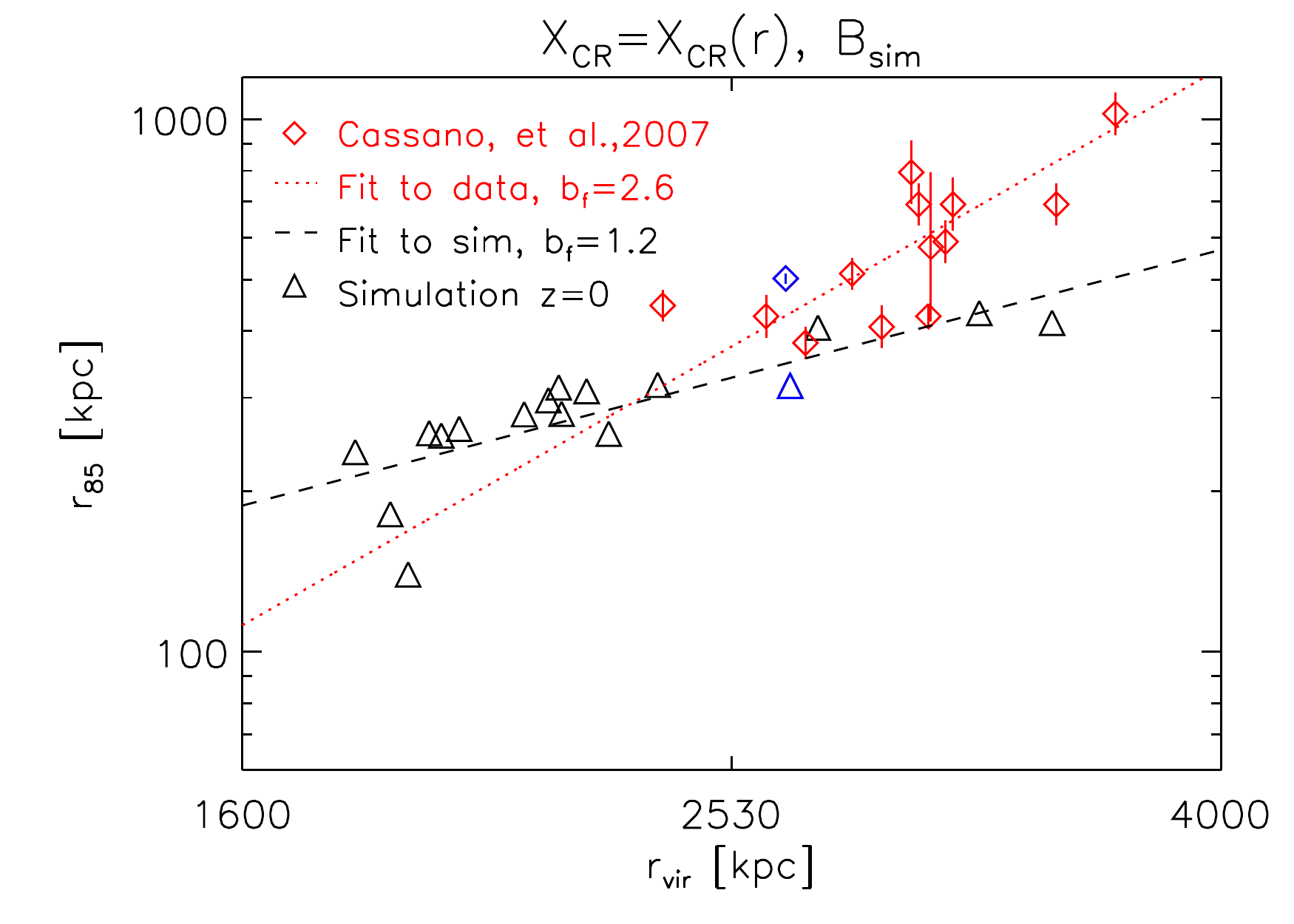}
\includegraphics[width=0.33\textwidth]{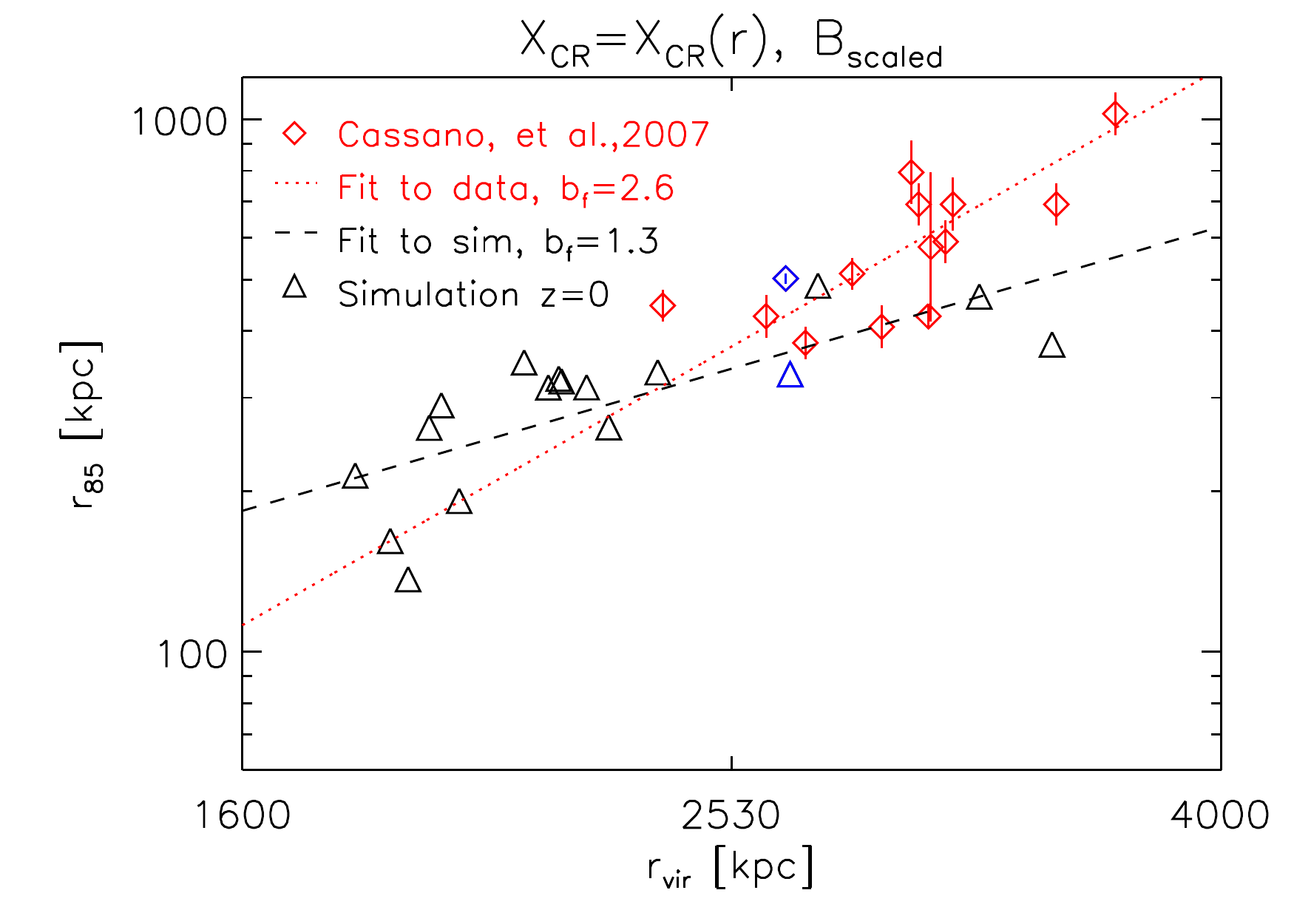}
\caption{Radius of the radio emitting region as function of the 
virial radius of the cluster. We plot, left to right, the correlation for all
  three models and include the best fit 
  ($\chi^2=$ 0.54, 0.29, 0.86, respectively) . We also
  plot observations of 14 galaxy clusters from
  \citep{2007MNRAS.378.1565C} and the best fit correlation. For both, 
simulations and observations the Coma cluster is marked in blue. For model 1 
we include the include the correlation at redshift 0.4.
}\label{size_scal}
\end{figure*}

As stated in Section \ref{scalrel}, observations of clusters with 
radio halos show that the size of halos scales not linearly with the 
cluster virial radius, 
suggesting that the non-thermal component in clusters is not self-similar.

In figure \ref{size_scal} we plot the radius containing 85\% of the clusters 
emission over virial radius for all three models. We include data
from \citet{2007MNRAS.378.1565C} and the fits to the simulated
clusters-distribution obtained for models 1--3 at $z=0$.
We find a correlation between $R_{85}$ and $R_{\mathrm{vir}}$ 
for our simulated hadronic-halos.
Results suggest that self-similarity is preserved 
in the non-thermal components, as the increase of the halos's 
radius is roughly proportional to the virial radius of the hosting clusters;
the slopes of the correlations are $b_{\mathrm{f}}=$ 0.8, 1.2 and 1.3 for
models 1, 2 and 3, respectively.

\noindent
This is not in line with observations: the expected correlations 
are flatter than the observed one and we predict halos systematically 
smaller than the observed ones.
In contrast this is expected considering results reported in section \ref{rradprof} and
confirms that it is challenging to reproduce the extension of the 
observed radio halos with hadronic models, even by adopting a profile
of the magnetic field that is flatter than that from our MHD simulations 
and by assuming a flat profile of the spatial distribution of CRp (model 3).

\noindent
\citet{2010MNRAS.401...47D} have shown that matching the radio emission of the 
Coma halo at distance $\approx 0.2-0.3 \, R_{vir}$ with hadronic models
(by further inceasing the CRp energy content at larger radii, 
see also section \ref{rradprof}) would require the energy content 
of CRp to be roughly similar to the content of the thermal ICM at these distances (at least when 
constraints on the magnetic field from RM observations of the Coma cluster
are used).
The scalings found in figure \ref{size_scal} make the situation possibly
more challenging, because they show that even more energy in the form of 
CRp would be required 
in the case of more massive clusters where the differences, between $R_{85}$
of our hadronic-halos and that of the observed ones, are larger.

The observed clusters are in a range of redshifts, $z=0 -- 0.4$, while 
hadronic-halos are extracted from the simulations at $z=0$.
For this reason in figure \ref{size_scal}a we also plot 
the distribution of simulated clusters obtained at $z=0.4$ in the case 
of model 1 (marked green).
We find that the correlation between $R_{85}$--$R_{vir}$ does not show 
significant change in slope at higher redshift.

\subsubsection{The X-ray Luminosity vs. Radio power relation and the 
evolution of radio halos}

\begin{figure*}
\centering
\includegraphics[width=0.33\textwidth]{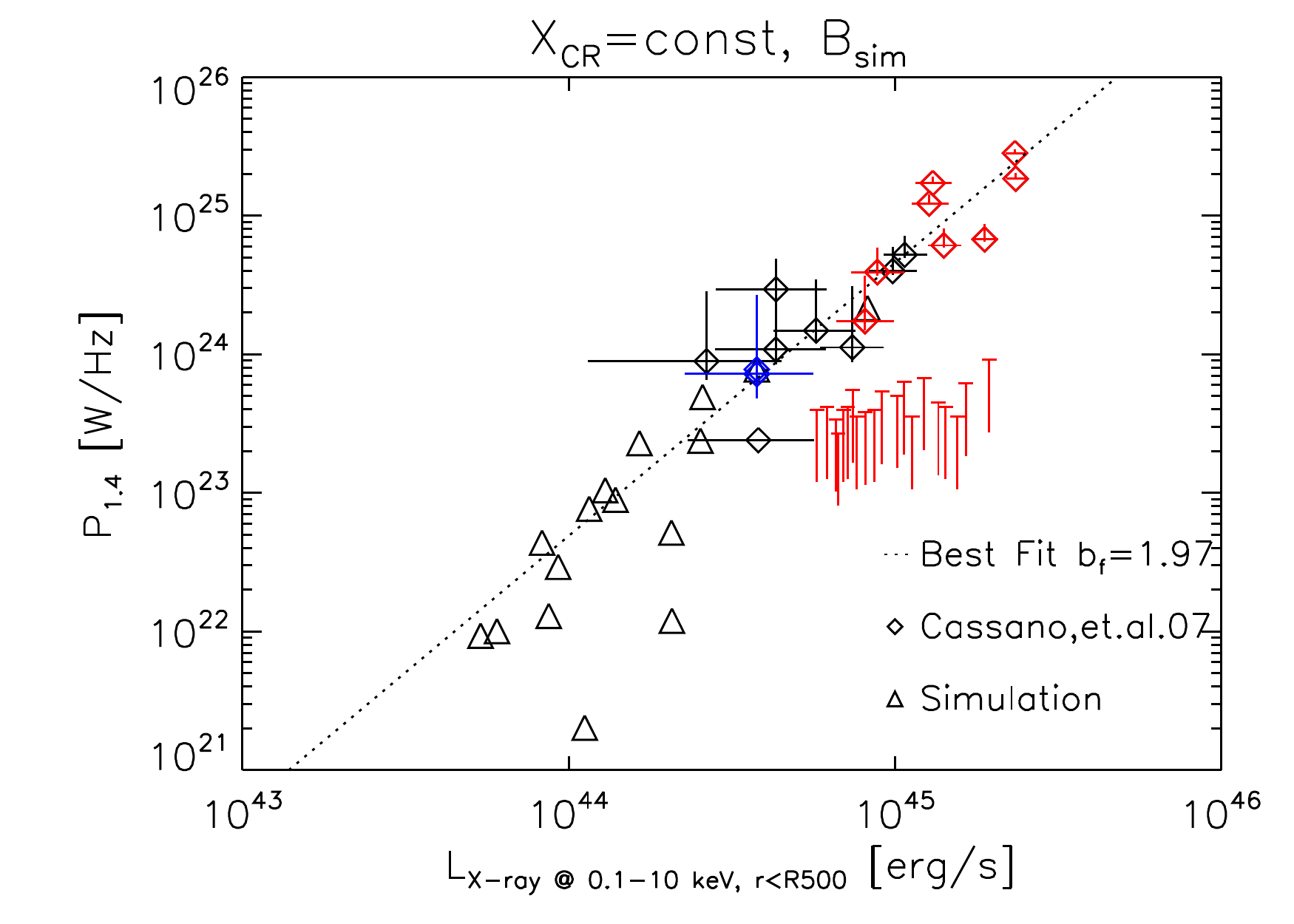}
\includegraphics[width=0.33\textwidth]{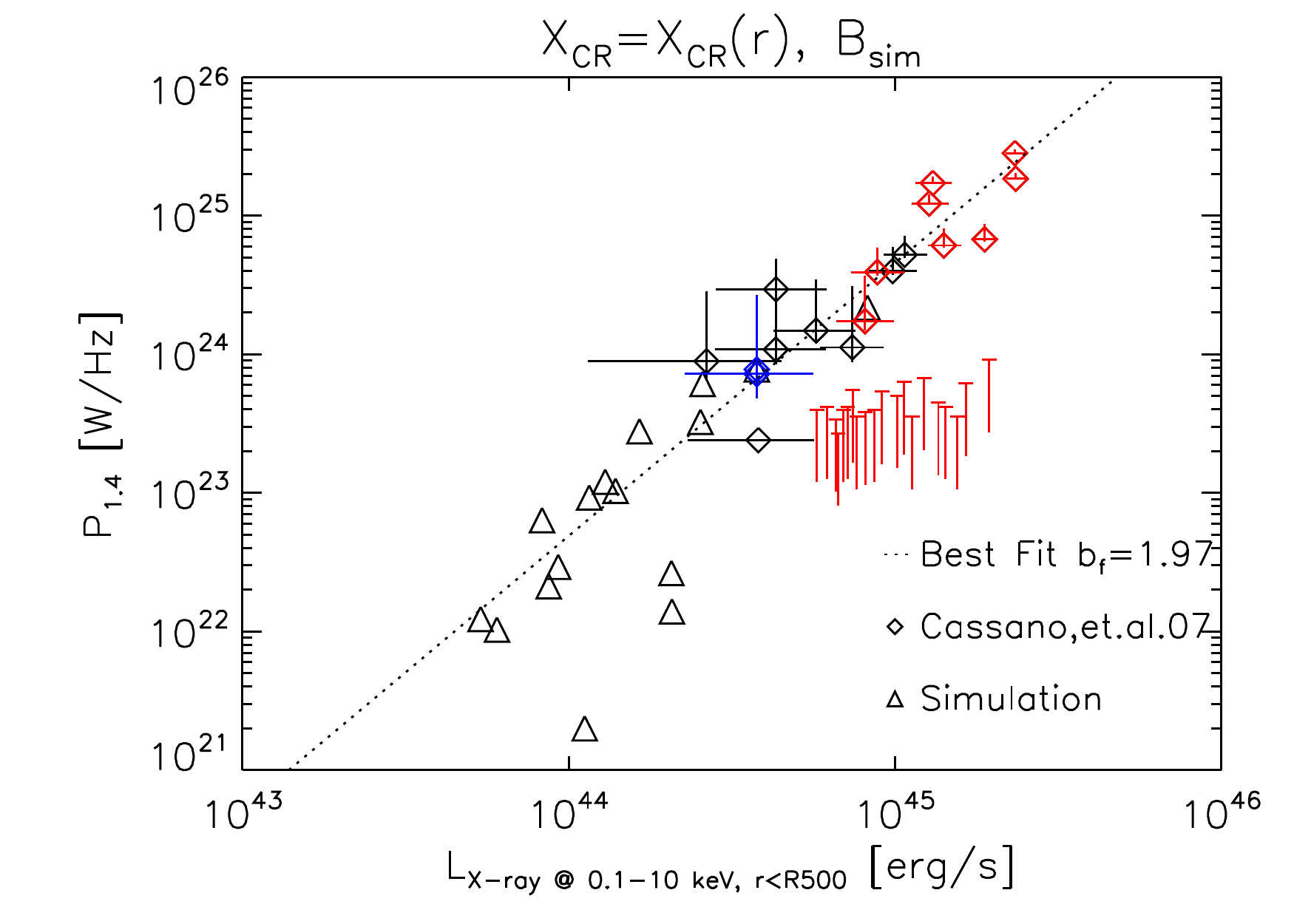}
\includegraphics[width=0.33\textwidth]{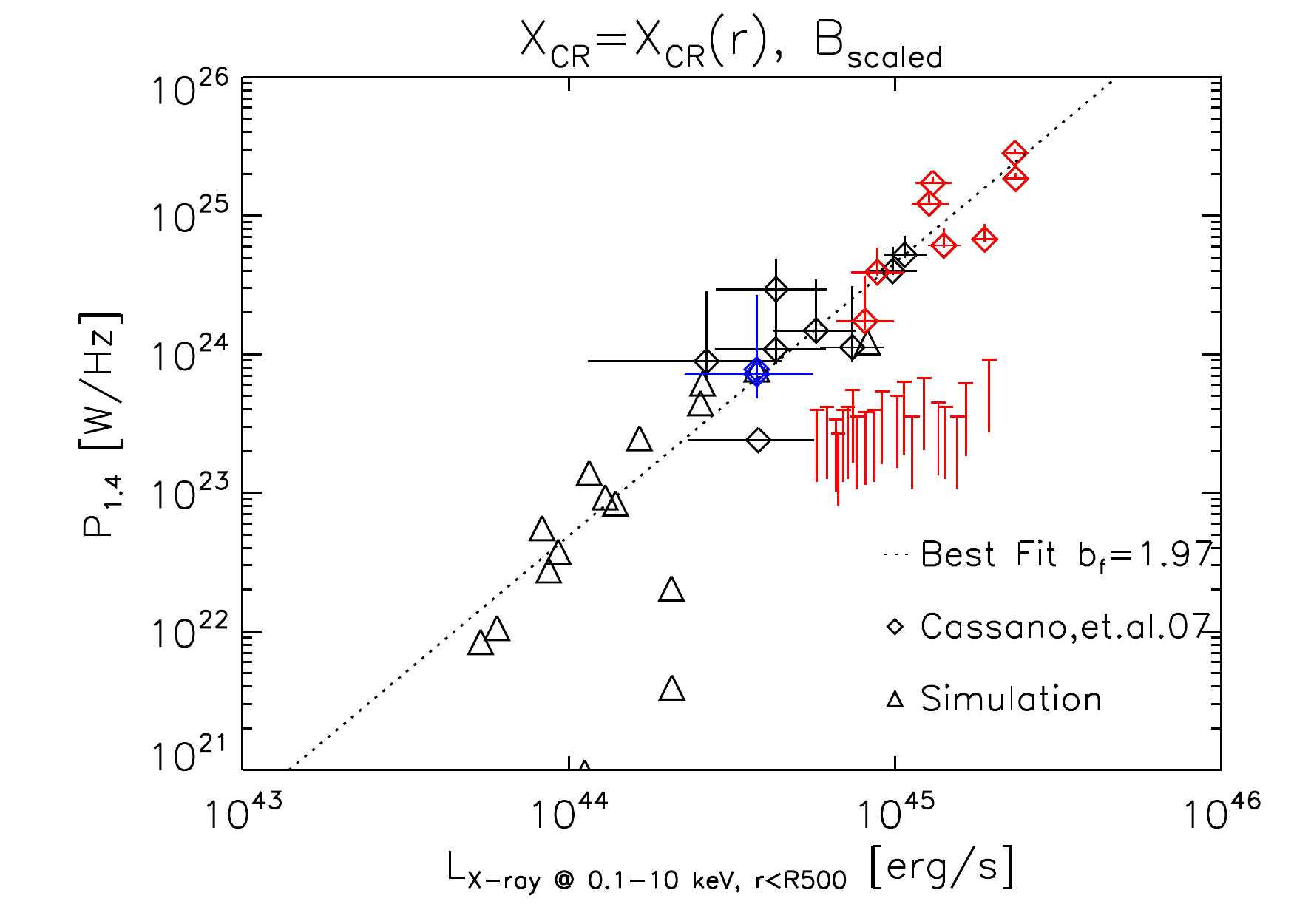}
\caption{Radio power per frequency at $1.4\,\mathrm{GHz}$ over X-ray
luminosity in erg/s from our simulated clusters using all three
models (triangles, ltr.: constant fraction, scaled fraction,scaled
fraction and scaled field). Observed scalings by
\citet{2007MNRAS.378.1565C} (diamonds, 
$z > 0.2$: red diamonds) and non-detections in red \citep{2007A&A...463..937V,2008A&A...484..327V}.
For both, simulations and observations the Coma cluster is marked
blue. }\label{r_x}
\end{figure*}

Radio halos follow a correlation between the monochromatic radio luminosity 
at 1.4 GHz, $P_{1.4}$, and the X-ray luminosity of the hosting clusters, 
$L_X$ \citep[e.g.][]{2000ApJ...544..686L,2002A&A...396...83E,
2003A&A...400..465B,2006MNRAS.369.1577C}.
Recent radio observations of a statistical sample of X-ray selected 
galaxy clusters, the ``GMRT radio halo survey'',  
\citep{2007A&A...463..937V,2008A&A...484..327V} 
allow to study the distribution of clusters in the $P_{1.4}-L_X$ 
diagram. These observations suggest that the distribution of clusters in 
the $P_{1.4}-L_X$ diagram is {\it bi-modal}: radio-halo clusters trace 
the $P_{1.4}-L_X$ correlation, while the majority of clusters are 
found ``radio quiet'' with the limits to their radio luminosities about 
10 times smaller than the radio luminosities of halos.

\noindent
In order to investigate the behaviour of our simulated clusters in the 
$P_{1.4}-L_X$ diagram, in Fig.\ref{r_x} we plot our simulated clusters 
together with observed clusters (from \citet{2008A&A...484..327V}).
The synthetic radio luminosities of our simulated clusters are scaled in order 
to have the simulated Coma cluster matching the observed one.\\ 

According to secondary models a correlation between radio luminosity and 
cluster X--ray luminosity (or temperature) is expected 
\citep[e.g.][]{2000A&A...362..151D,2001CoPhC.141...17M,2005JCAP...01..009D,
2008MNRAS.385.1211P}.
We qualitatively confirm these expectations and in all three models
find that the largest simulated clusters would naturally approach the 
observed correlation.
In all three models the smallest systems are significantly more scattered in radio power
than the largest clusters. This is expected as the magnetic field in the
central regions of our simulated clusters (where most of the synchrotron 
emission is generated) is found to be tighly correlated with cluster thermal 
properties only in the case of massive clusters, while only a steep trend 
is found for smaller systems \citep[figure 9 in][]{2008arXiv0808.0919D}.

The most relevant difference with respect to observations is that
according to all three models (as for every secondary model)
the synchrotron luminosity of the massive simulated clusters 
is equivalent to that of typical radio halos, at least if the radio
luminosity of the simulated Coma cluster is normalised
to that of the real Coma halo.
This is inconsistent with observations which, on the other hand, found 
radio halos in only about $1/3$ of massive clusters.

\noindent
Most important, no radio {\it bi--modality} is expected in our simulated 
secondary--halos. Hadronic halos in simulated
massive systems would follow a tight correlation, 
while those in less massive systems would be more broadly distributed.

\noindent
One possibility to reconcile the hadronic scenario with the 
observed halo--merger connection and with the {\it bi--modal} distribution 
of clusters in the $P_{1.4}-L_X$ diagram is to admit that the observed 
{\it bi--modality} is driven by the amplification and dissipation of the 
magnetic field in the merging and post--merging phase, respectively 
\citep{2007MNRAS.378..245B,2008MNRAS.385.1211P,2009JCAP...09..024K}.
However, Burnetti et al.(2009) have shown that the degree of
amplification/dissipation of the magnetic field and 
the time--scale of this process that would be necessary to
explain observations are difficult to reconcile with the observed 
properties of magnetic fields in the ICM (namely the field intensity
and coherence scales from Rotation Measurements), and appear also 
disfavoured by energetic arguments.

\begin{figure}
\centering
\includegraphics[width=0.5\textwidth]{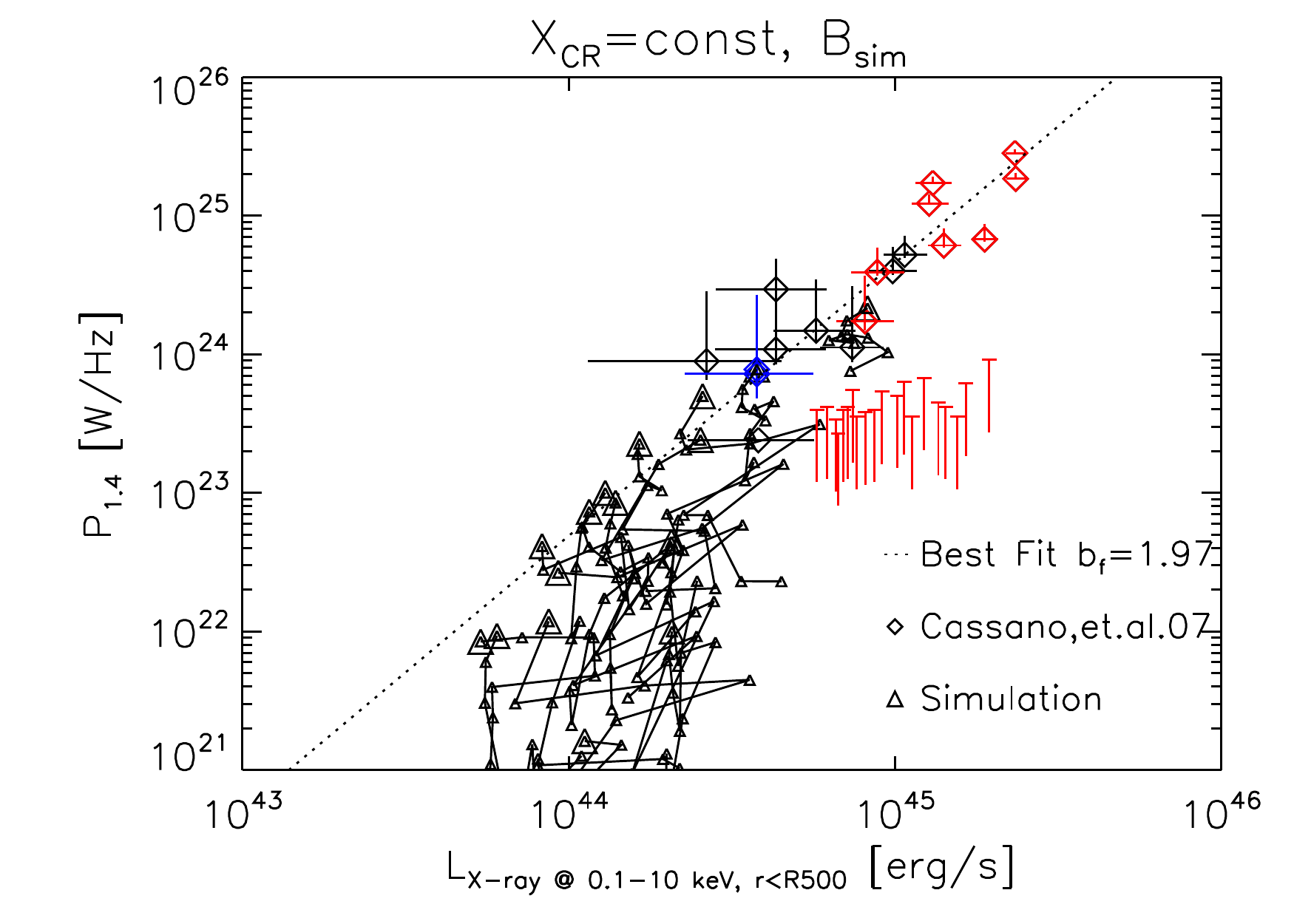}
\caption{Radio power per frequency at $1.4\,\mathrm{GHz}$ over X-ray
  luminosity in erg/s from our simulated clusters using the constant
  model. Observed scalings by \citet{2007MNRAS.378.1565C} (diamonds, $z > 0.2$: red diamonds)
  and non-detections in red \citep{2007A&A...463..937V,2008A&A...484..327V}.
  We also include the time evolution of clusters for $z
  < 0.48$ (black lines)}\label{r_x_evo}
\end{figure}

\noindent
The amplification of the magnetic field during cluster mergers is 
followed by our MHD cosmological simulations that offer a complementary 
approach to highlight this issue.
To show the effect of magnetic field evolution and
investigate the {\it bimodality} we plot the time
evolution of the simulated radio haloes in figure \ref{r_x_evo}.
Shown is the radio luminosity over X-ray luminosity of the simulated 
sample for redshift $z < 0.48$ (big triangles:
z=0, small triangles and black line: earlier redshifts); also 
we show the radio power vers. temperature correlation and evolution 
in the Appendix.
No hint of a {\it bi--modality} is found, simply because 
magnetic field amplification in massive systems is a gradual process 
that happens in a time--scale comparable with the life--time of clusters 
themselves. 
The evolution of magnetic field is reflected in the broad/scattered 
distribution of 
clusters in the $P_{1.4}-L_X$ diagram, especially in the case of smaller 
systems.
These smallest clusters approach the lower end of the correlation
with rather large scatter because the magnetic field is not saturated 
yet in their central regions and even smaller mergers 
yield a significant field amplification resulting in an increased 
radio luminosity.
Although one would expect this behavior to be slightly dependent on numerical 
resolution, increasing the numerical resolution would make the situation 
even more stringent. Resolving smaller gas motions leads to increase the 
amplification of the magnetic field, specially at early times. 
Having a larger magnetic field the changes in the magnetic field due to
merger activity will be suppressed as the magnetic field is already
closer to saturation effects.
Therefore we would expect the clusters (especially the smaller, less
resolved systems) to evolve even more along the correlation, and
the spread around it would be further decreased.

\section{$\gamma$-ray emission from simulated clusters}

\begin{figure}
\centering
\includegraphics[width=0.5\textwidth]{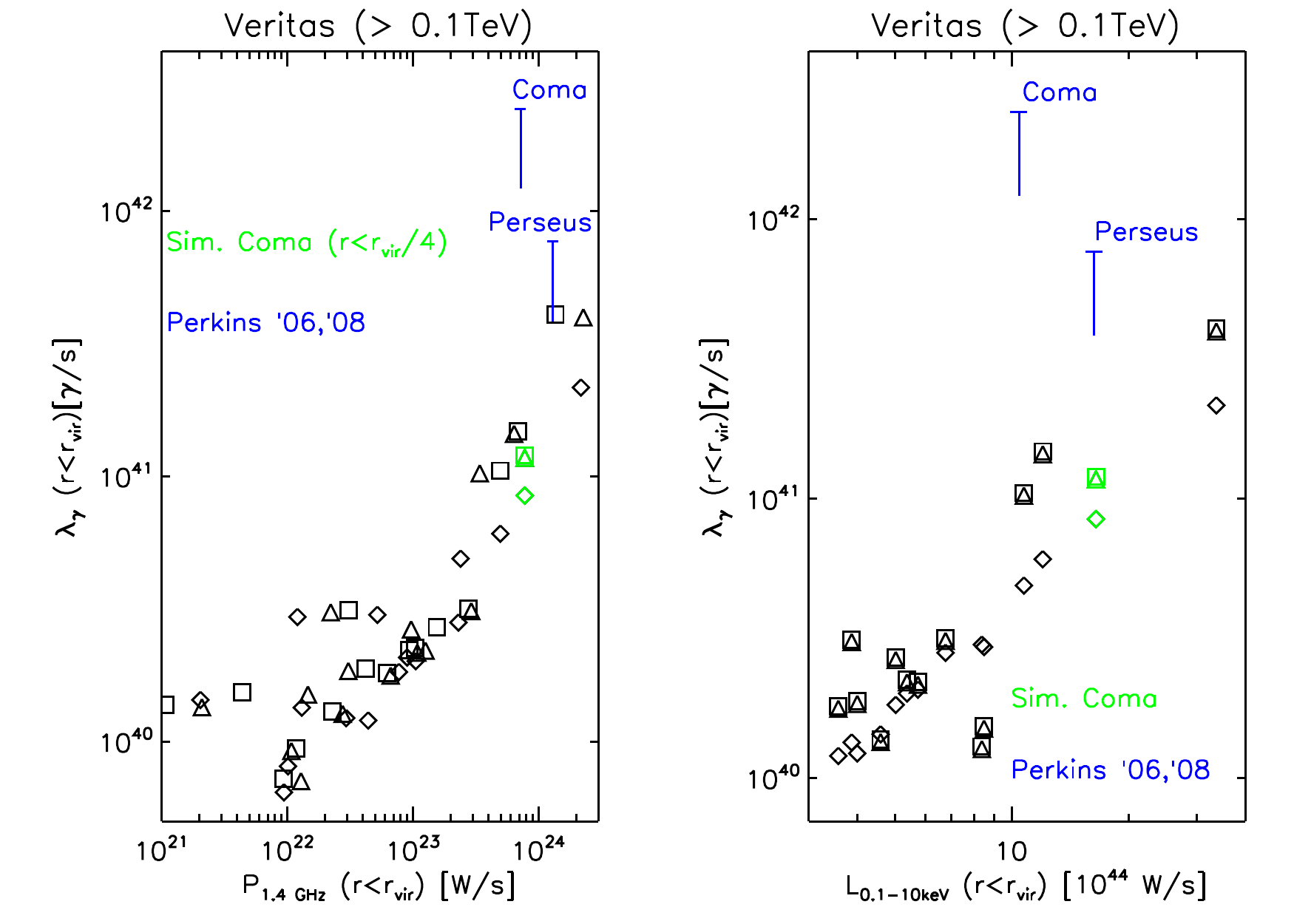}
\caption{$\gamma$-ray luminosity ($E>0.1\,$Tev, e.g. VERITAS) over radio luminosity (at 1.4 GHz,left) and bolometric X-Ray luminosity of all cluster at redshift z=0.We plotted the three 
models as different symbols (diamonds - model 1, triangles - model 2, boxes - model 3) and include upper limits from 
\citet{2008AIPC.1085..569P,2006ApJ...644..148P} in blue.  The Coma cluster is marked green.}\label{veritas}
\end{figure}

\begin{figure}
\centering
\includegraphics[width=0.5\textwidth]{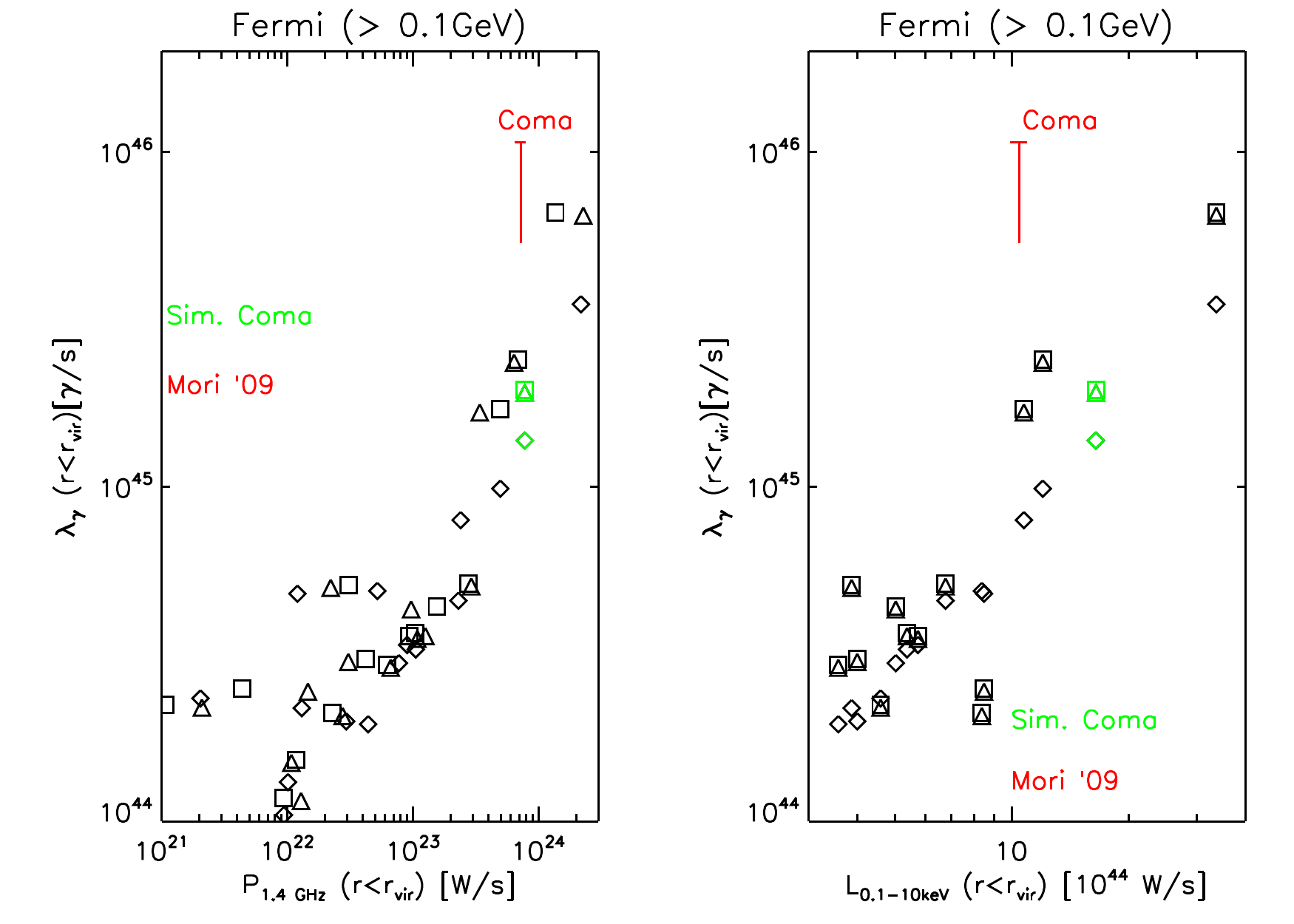}
\caption{$\gamma$-ray luminosity ($E>0.1\,$Gev, e.g. FERMI, EGRET) over radio luminosity (at 1.4 GHz,left) and bolometric X-Ray luminosity of all cluster at redshift z=0.We plotted the three 
models as different symbols (diamonds - model 1, triangles - model 2, boxes - model 3) . We add preliminary results from the first year FERMI data \citep{2009arXiv0912.3346M} in red. The Coma cluster is marked green.  }\label{fermi}
\end{figure}

We use the formalism in section \ref{form_gamma} to compute 
predictions for the $\gamma$-ray
luminosity of the simulated clusters according to all three models. 
Of interest are the two energy bands of Cherenkov telescopes, 
$E > 0.1 \, \mathrm{TeV}$, and 
FERMI/EGRET telescopes, $E > 0.1 \, \mathrm{GeV}$. \\
In table \ref{limits_veritas} (\ref{limits_fermi}) we present fluxes
for the simulated sample in the VERITAS (FERMI) energy band. 
None of the clusters is in the observable range of the 
VERITAS experiment, however the largest ones (0, Coma, Virgo, Perseus, 
Centaurus) have a chance to be detected by FERMI in next
years, at least for models with 
radially increasing CRp normalisation (model 2 and 3).
\\

In the left panel of figures \ref{veritas} and \ref{fermi} we plot the simulated clusters
in $\gamma$-ray vers. radio luminosity at 1.4 GHz for the 3 models.
Following 
the previous Section we normalise the radio luminosity of the 
sample so the simulated Coma cluster fits the observed emission from 
the Coma halo. We also report available limits from VERITAS observations 
\citep{2008AIPC.1085..569P,2006ApJ...644..148P} (figure \ref{veritas}). 
Further we include (figure \ref{fermi}) newest 
preliminary results from the FERMI experiment \citep{2009arXiv0912.3346M}
on the upper limit of the $\gamma$-ray flux from the Coma cluster.
A scaling between $\gamma$-ray and radio luminosity is expected 
for the 3 models because p--p collisions generate both
secondary electrons and neutral pions, the scaling is sub--linear 
since the radio luminosity is further boosted by the increase of the 
magnetic field in more massive clusters.\\

In the right panel of figure \ref{veritas} and \ref{fermi} we plot the emission of our clusters in $\gamma$-ray vers. X-ray luminosity for the 3 models.
Also in this case we show two $\gamma$-ray upper limits from 
VERITAS \citep{1992A&AS...95..129P} and preliminary FERMI \citep{2009arXiv0912.3346M} 
observations respectively. 
A quasi--linear correlation is predicted, and is found less scattered
than that between $\gamma$-ray and radio luminosities since it
compares two purely thermal quantities (also in the case of models
2 and 3 CRp are scaled with thermal energy according with the profile
in Figure \ref{img_xcr}). \par
\noindent
The FERMI results  are still consistent with the models for CRp presented in this paper. 
As shown by \citet{2010MNRAS.401...47D} the expected $\gamma$-ray flux increases substantially 
if we consider a spatial CR proton distribution so the radio emission actually fits the observations in morphology. 
We therefore expect a rejection or confirmation of our models in the  very near future.

\begin{table}
\centering
 \begin{tabular}{c|c|c|c} \hline
Cluster & M1 & M2 & M3 \\\hline
0 		 		& $3.2 \times 10^{-13}$ & $5.8 \times 10^{-13}$ & $6.0 \times 10^{-13}$ \\
HYDRA  		& $6.9 \times 10^{-14}$ & $3.3 \times 10^{-14}$ & $3.4 \times 10^{-14}$ \\
2 		 		& $2.4 \times 10^{-14}$ & $1.0 \times 10^{-14}$ & $1.0 \times 10^{-14}$ \\
3 		 		& $1.7 \times 10^{-14}$ & $2.5 \times 10^{-14}$ & $2.5 \times 10^{-14}$ \\
4 		 		& $9.9 \times 10^{-14}$ & $2.0 \times 10^{-13}$ & $2.1 \times 10^{-13}$ \\
5 		 		& $1.7 \times 10^{-14}$ & $2.6 \times 10^{-14}$ & $2.6 \times 10^{-14}$ \\
COMA   		& $6.7 \times 10^{-14}$ & $9.4 \times 10^{-14}$ & $9.6 \times 10^{-14}$ \\
7 		 		& $8.6 \times 10^{-15}$ & $9.8 \times 10^{-15}$ & $1.0 \times 10^{-14}$ \\
8 		 		& $1.8 \times 10^{-14}$ & $2.0 \times 10^{-14}$ & $2.1 \times 10^{-14}$ \\
VIRGO  		& $2.9 \times 10^{-13}$ & $6.7 \times 10^{-13}$ & $6.9 \times 10^{-13}$ \\
A3627  		& $3.2 \times 10^{-14}$ & $3.3 \times 10^{-14}$ & $3.4 \times 10^{-14}$ \\
11 	 		& $1.4 \times 10^{-14}$ & $2.1 \times 10^{-14}$ & $2.1 \times 10^{-14}$ \\
12 	 		& $4.5 \times 10^{-14}$ & $4.2 \times 10^{-14}$ & $4.3 \times 10^{-14}$ \\
PERSEUS		& $9.7 \times 10^{-14}$ & $2.3 \times 10^{-13}$ & $2.3 \times 10^{-13}$ \\
CENTAURUS	& $5.3 \times 10^{-14}$ & $5.8 \times 10^{-14}$ & $5.9 \times 10^{-14}$ \\
15 			& $5.0 \times 10^{-15}$ & $5.5 \times 10^{-15}$ & $5.6 \times 10^{-15}$ \\

\hline
\end{tabular}
\caption{Fluxes ($\gamma/\mathrm{cm}^{2}/\mathrm{s}$) in the VERITAS energy 
range ($E>100\,$ GeV) from $1\, R_{\mathrm{vir}}$ for all Clusters from the 
simulations using the three different models.} \label{limits_veritas}
\end{table}

\begin{table}
\centering
\begin{tabular}{c|c|c|c} \hline
Cluster & M1 & M2 & M3 \\\hline
0 				& $5.2 \times 10^{-9}$  & $9.5 \times 10^{-9}$  & $9.7 \times 10^{-9}$  \\
HYDRA  		& $1.1 \times 10^{-9}$  & $5.5 \times 10^{-10}$ & $5.6 \times 10^{-10}$ \\
2 				& $3.9 \times 10^{-10}$ & $1.6 \times 10^{-10}$ & $1.7 \times 10^{-10}$ \\
3 				& $2.8 \times 10^{-10}$ & $4.1 \times 10^{-10}$ & $4.1 \times 10^{-10}$ \\
4 				& $1.6 \times 10^{-9}$  & $3.4 \times 10^{-9}$  & $3.4 \times 10^{-9}$  \\
5 				& $2.8 \times 10^{-10}$ & $4.2 \times 10^{-10}$ & $4.3 \times 10^{-10}$ \\
COMA  		& $1.1 \times 10^{-9}$  & $1.5 \times 10^{-9}$  & $1.5 \times 10^{-9}$  \\
7 				& $1.4 \times 10^{-10}$ & $1.6 \times 10^{-10}$ & $1.6 \times 10^{-10}$ \\
8 				& $3.0 \times 10^{-10}$ & $3.3 \times 10^{-10}$ & $3.4 \times 10^{-10}$ \\
VIRGO  		& $4.7 \times 10^{-9}$  & $1.0 \times 10^{-8}$  & $1.1 \times 10^{-8}$  \\
A3627  		& $5.2 \times 10^{-10}$ & $5.4 \times 10^{-10}$ & $5.5 \times 10^{-10}$ \\
11 			& $2.3 \times 10^{-10}$ & $3.4 \times 10^{-10}$ & $3.5 \times 10^{-10}$ \\
12 			& $7.4 \times 10^{-10}$ & $6.9 \times 10^{-10}$ & $7.0 \times 10^{-10}$ \\
PERSEUS  	& $1.5 \times 10^{-9}$  & $3.7 \times 10^{-9}$  & $3.8 \times 10^{-9}$  \\
CENTAURUS  	& $8.7 \times 10^{-10}$ & $9.5 \times 10^{-10}$ & $9.7 \times 10^{-10}$ \\
15 			& $8.1 \times 10^{-11}$ & $9.0 \times 10^{-11}$ & $9.2 \times 10^{-11}$ \\
\hline
\end{tabular}
\caption{Fluxes ($\gamma/\mathrm{cm}^{2}/\mathrm{s}$) in the EGRET/FERMI 
energy range ($E>0.1\,$ TeV) from $1\,R_{vir}$
for all Clusters from the simulations using the three different models.} 
\label{limits_fermi}
\end{table}

\section{Conclusions}\label{conclusions}

We use a constrained, cosmological MHD SPH simulation with a semianalytic 
model for galactic magnetic outflows, to obtain a sample of 16 galaxy 
clusters with thermal properties 
similar to clusters in the Local Universe. 
Further we assume 3 different models for secondary cosmic rays motivated 
by simulations \citep{2007MNRAS.378..385P} using the proper high 
energy approximation for the pion cross-section. 
In the first model we assume a constant CRp normalisation relative to 
the thermal density and the simulated magnetic field. 
In the second model we keep the magnetic field and introduce 
a radius dependent CRp normalisation infered from non-radiative simulations 
by \citet{2007MNRAS.378..385P}. 
The third model uses the same CRs and flatten the simulated magnetic 
field to be $B \propto \sqrt{\rho}$.\\

\par 
Although our simulations do not include a (internally) self-consistent
treatment of CRp as done in other simulations \citep[e.g.][]{2007MNRAS.378..385P},
contrary to previous work they allow to properly simulate the properties 
of the magnetic field in the ICM which is important for modeling the
cluster-scale synchrotron emission.

\par
For the first time, we carry out a detailed comparison between the observed 
properties of giant radio haloes and those of simulated halos according to
secondary models and under different assumptions for the spatial distribution
of CRp.

\par
In an earlier paper we presented a detailed comparison of the simulated 
Coma cluster \citep{2010MNRAS.401...47D}. 
In this work we focus on global sample properties and compare with 
recent observations. \\

In particular, as a first step, we show that :

\begin{itemize}
\item The radial profiles of Faraday rotation of the median of our 
cluster sample is in line with that obtained from a number of observations 
of different clusters. This confirms that the properties of the clusters's
magnetic fields in our simulations are rather similar to the observed ones.

\item The normalised radial profiles of the radio emission at 1.4 GHz of 
our simulated hadronic-halos show a deficit at radii $\geq 0.1 r_{\mathrm{vir}}$ 
with respect to the synchrotron profiles observed for a sample of well 
studied radio halos. 
This is in line with previous claims based on semi-analytic calculations
in the context of the hadronic model \citep{2004MNRAS.350.1174B}. 
In addition our results show that, even by
assuming a flat profile for both the magnetic field and CRp spatial
distributions (Model 3), secondary electrons may account for less than
about 10-15 \% at radii 0.15 - 0.3 $r_{\mathrm{vir}}$.

\item A Point to Point comparison of radio vs. X-ray emission, obtained for 
the 4 largest clusters of our sample, confirms the results obtained for 
the profiles showing that the radial distribution of radio emission is 
too steep.
All three models do not fit the observations over the whole range. 
Furthermore an excess in radio emission of the innermost patches 
suggests that haloes from the simulation are to centrally peaked.

\end{itemize}

As a second step we compare scaling relations obtained for the 
hadronic-halos in our simulated cluster sample with those given
in \citep{2007MNRAS.378.1565C} that are obtained
from a sample of observed radio halos.
We find that :

\begin{itemize}

\item The geometrical correlation between the radius of radio halos
and the cluster-mass contained within this radius is well reproduced
by all three models.
Due to the expected self-similarity of cluster in thermal properties 
this result implies that -at least- the simulated and observed clusters
share similar physical properties.

\item A quasi self-similar behaviour is found for the non-thermal properties 
of our simulated clusters. In particular the radius of our hadronic halos 
is found to scale (approximately) with the virial radius of the simulated
clusters. This is contrary to observations that found a steeper correlation
between halo-radius and virial radius of the hosting clusters and 
implies that our simulated halos are systematically smaller than the
observed ones.

\item A correlation between the monochromatic luminosity of our hadronic
halos and the X-ray luminosity of the simulated hosting-clusters is found.
As soon as the population of our hadronic halos is normalised, by scaling 
the radio luminosity of the simulated Coma halo with that of the observed
one, the correlation is similar to that observed for radio halos.
However, since at this point all the simulated clusters show radio
emission at the level of the observed halos, we find that the cluster
radio {\it bi-modality}, observed for X-ray selected clusters, cannot
be reproduced.
A radio {\it bi-modality} in the radio -- X-ray diagram would require
a fast ($<$ Gyr) evolution of the radio luminosity in connection with 
cluster mergers, on the other hand, we find that the time evolution of 
our simulated massive-clusters in this diagram happens on cosmological, 
long, time-scales.
Finally, we show that once the radio -- X-ray correlation is approached
at low X-ray luminosities, clusters follow the correlation closely
due to the saturation of the magnetic field.

\end{itemize}

As a final point we calculate the $\gamma$-ray emission from our simulated
clusters once the radio luminosity of the simulated Coma halo
is anchored to that of the observed one (essentially by scaling the number
density of CRp in simulated clusters).
We find that :

\begin{itemize}

\item The $\gamma$-ray emission expected from our simulated clusters 
is well below the sensitivity of present Cherenkov Arrays, e.g.
the VERITAS experiment, for all the adopted models for hadronic halos.
The $\gamma$-ray fluxes at $> 100$ MeV expected from our
simulated clusters would allow for a marginal detection by the FERMI 
telescope in next years, at least by assuming Models 2 and 3.

\item The integrated $\gamma$-ray flux from our simulated clusters
is expected to scale with their radio emission, although with rather
large scatter.
On the other hand, a tight correlation is found between $\gamma$-ray
and X-ray fluxes, in which case (due to the scalings between thermal and
non-thermal CRp in Models 1--3) the correlation is essentially driven
by the density of the thermal gas in our simulated clusters.

\end{itemize}

Considering all these results, as well as the outcome from our earlier work 
on the Coma cluster \citep{2010MNRAS.401...47D}, we conclude that hadronic models 
alone are not able to explain the observed properties of giant radio haloes, 
in terms of their radial extension, observed clusters's radio 
{\it bi-modality}, scaling relations and spectral properties. Therefore we conclude 
that radio emission observed in galaxy clusters in form of giant radio halos are a
powerful tool to infer the amount of cosmic ray protons within galaxy cluster,
confirming previous results that the energy content of cosmic ray protons in clusters
can not exceed percent level \citep[e.g.][]{2007ApJ...670L...5B,2008MNRAS.388.1062C}. 

The problem of halo's extension could be alleviated in 
{\it extended} hadronic models, where the contribution from
primary (shock accelerated) electrons at the cluster outskirts is combined
in the simulations with that from secondary electrons (\citet{2008MNRAS.385.1211P}). 
However, as shown in Sect. \ref{rradprof}, the problem of the halo profiles 
arises at distances 0.1 -- 0.3 $R_\mathrm{vir}$ from cluster centers where, based on
the same simulations, the contribution to the diffuse synchrotron 
emission from these shock accelerated electrons is not yet dominant.
Further we would like to note that the detailed morphology of the radio emission caused
by electrons injected at shocks \citep[e.g.][]{1999ApJ...518..603R}
\footnote{These are base on Eularian simulation, which are most appropriate to describe
the morphology of shocks outside the cluster core.} 
will be significantly different from the morphology of giant radio halos and 
therefore is in general unlikely able to mimic large scale, cluster 
centric radio emission.
Thus additional physical processes are required to explain observations.

The problem of the radio {\it bi-modality} in massive clusters can be alleviated
by assuming that MHD turbulence (and the rms magnetic field) decays as 
soon as clusters approach a relaxed state after a major merger \citep{2009A&A...507..661B,2009JCAP...09..024K}.
In this case it might also be thought that CRp in these relaxed clusters
would undergo less scattering on magnetic field irregularities escaping
from the cluster volume and reducing the source of secondary electrons.
All these processes are not properly included in our MHD cosmological
simulations.
However, \citet{2009A&A...507..661B} have shown that the fast decay of the rms 
field in galaxy clusters, that is necessary to explain the observed 
{\it bi-modality}, would imply a {\it rather unphysical} situation where 
the magnetic field power spectrum peaks at the smaller scales, and leads 
to the consequence that an {\it extremely} large flux of energy in clusters 
goes into magnetic field amplification/dissipation.
In addition, because turbulent cascade should start to decay on largest 
scales, which are resolved by Faraday rotation measurements,  
this scenario would predict a {\it bi-modality} in rotation measures 
and depolarisation in largest clusters, which is not observed (Govoni et al. 2010).\\

The discovery of giant radio halos with {\it very} steep spectrum
($\alpha \sim 1.8-2$) in merging clusters proves that {\it inefficient} 
particle acceleration mechanisms are responsible for the origin of these 
sources and, based on simple energy arguments, disfavours hadronic models 
\citep{2008Natur.455..944B}.
Observations of the best studied halo, in the Coma cluster, have shown 
that its spectrum steepens at higher frequencies \citep{2003A&A...397...53T}, due to the
competition between energy losses and acceleration of the emitting
particles, in which case the particle acceleration time-scale would
be of the order of 0.1 Gyr.
More recently we have shown that the steepening cannot be a result 
of the inverse Compton (SZ) decrement \citep{2010MNRAS.401...47D}, 
confirming these previous finding and supporting a scenario of
{\it inefficient} particle acceleration mechanisms at the origin of the halo.

\noindent
Particle acceleration due to micro-turbulence in merging galaxy 
clusters has been proposed for the origin of radio halos \citep[e.g.][]{2001MNRAS.320..365B,2001ApJ...557..560P}. In this case {\it gentle} particle acceleration
mechanisms would generate radio halos in connection with (massive) cluster 
mergers, while the radio emission would decay as soon as clusters approach 
a relaxed state, due to dissipation of (at least a fraction of) this 
turbulence and the fast electron radiative cooling.

It will therefore be necessary to consider these processes in future simulations
used to study radio halos. That includes an estimation of the locally merger injected 
turbulence as well as a more detailed description of CR electron spectra.

\section{Acknowledgements}
J.~D.~kindly acknowledges the support of ESF/Astrosim Exchange grant 2065 and thanks the INAF/IRA in Bologna 
for the hospitality. K.~D.~acknowledges the supported by the DFG Priority Programme 1177.

\bibliographystyle{mn2e} \bibliography{master}

\appendix
\section{Additional Scaling Relations}

Here we provide additional scaling relations to substantiate the scalings shown before.

\subsection{X-Ray luminosity - Temperature relation}
\begin{figure}
   \centering
	\includegraphics[width=0.5\textwidth]{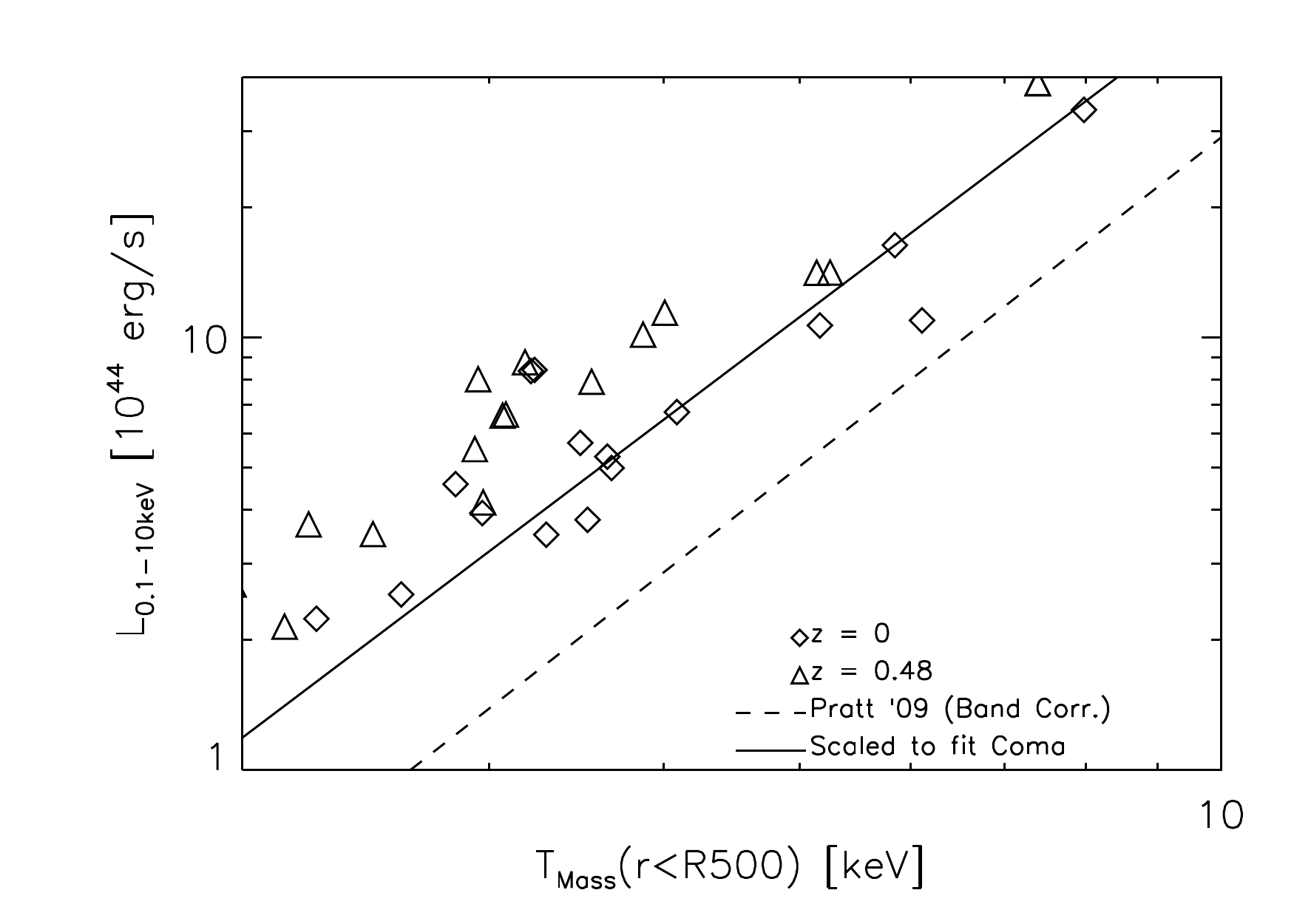}
	\caption{Bolometric X-Ray luminosity over temperature of all cluster at redshift z=0 (diamonds) and z=0.48 (triangles). We also plot the observed correlation from \citet{2008arXiv0809.3784P}.}\label{lx_t}
\end{figure}
In figure \ref{lx_t} we show bolometric X-ray luminosity over mass weighted temperature for the simulated cluster sample at redshift zero (diamonds) and 0.48 (triangles). We include band-corrected the observed relation from \citet{2008arXiv0809.3784P} (broken line) and a the same correlation normalised to the simulated Coma cluster (black line).  It is well known that non radiative simulations tend to over predict the x-Ray luminosity by a fair amount (see for example \citep{2006MNRAS.367.1641B}). In addition the effect of K-correction can be seen in comparison with the cluster at high redshift. \\
 Still, after re-normalizing the x-Ray luminosity there is a reasonably good agreement of with observations
We therefore conclude, that our simulated clusters sample shows similar thermal properties than the observations. 

\subsection{Radio luminosity - Temperature relation}

\begin{figure}
   \centering
	\includegraphics[width=0.5\textwidth]{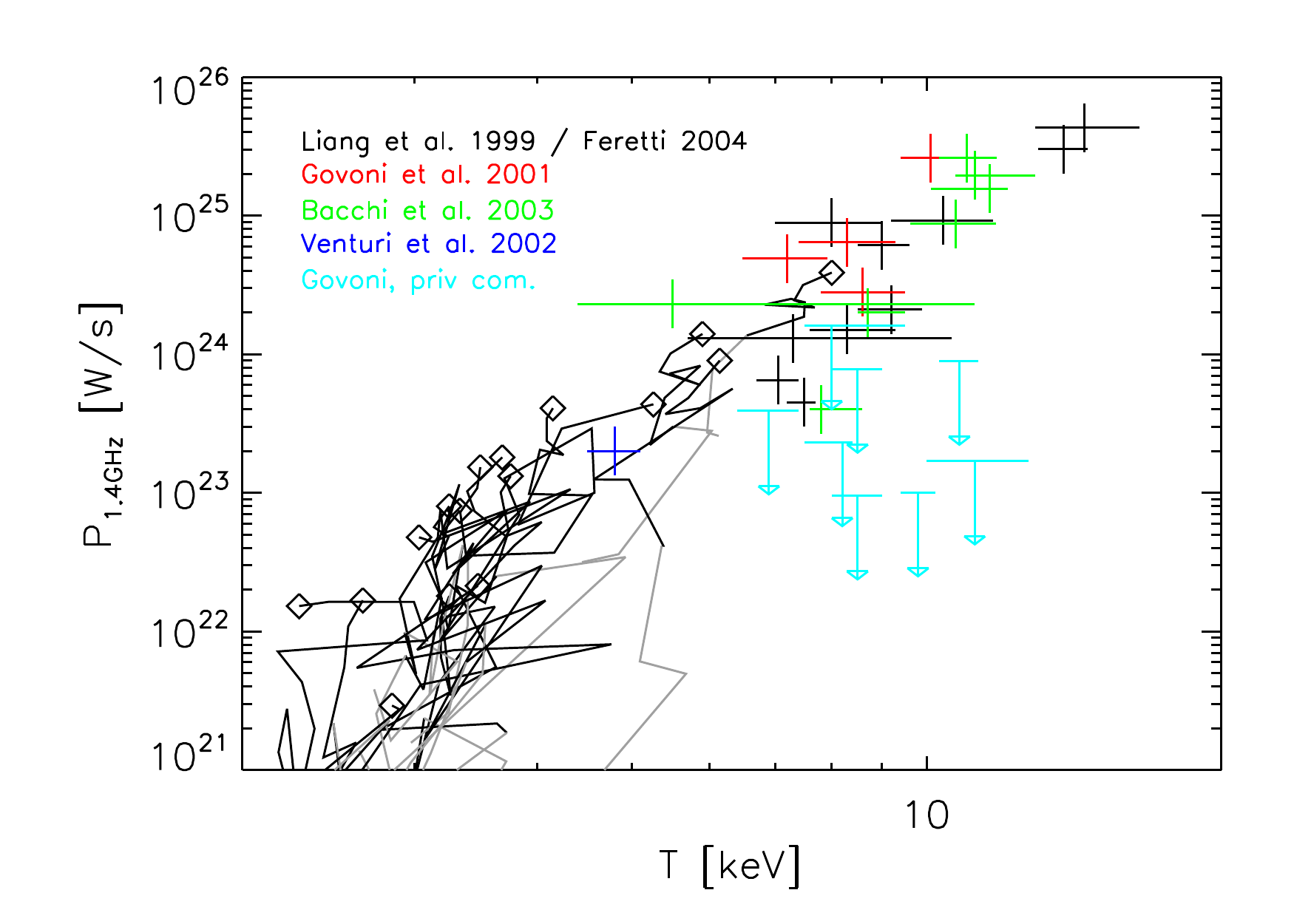}
	\caption{Radio Power at $1.4\,\mathrm{Ghz}$ over mass weighted temperature inside $0.1 r_{\mathrm{vir}}$ for all clusters. We include various observations from the literature. The evolution of the simulated clusters is plotted as lines in black (gray) for $z<0.48$ ($z>0.48$), while the points correspond to $z=0$.}\label{timevo}
\end{figure}
In figure \ref{timevo} we present the scaling of radio power at 1.4 GHz over cluster temperature as obtained from e.g. X-ray observations for constant CRp scaling (model 1). We include the evolution of the emission with time  as lines in black (gray) for $z<0.48$ ($z>0.48$). Further we plot a number of recent observations.
The sample follows the correlation closely at redshift zero, and only two clusters show significant deviation from the correlation at temperature larger than 5 keV, and only at high redshift. \\
In simulations, temperature is a less merger sensitive mass estimator compared to the X-ray luminosity, in terms of cluster mergers. We therefore conclude, that the bimodality observed in large clusters is not a result of biased mass estimation. 

\label{lastpage}
\end{document}